\documentclass[aps,prd,amssymb,showpacs,floatfix,preprint,nofootinbib]{revtex4-1}
\usepackage{amsmath,amsfonts,graphics,epsfig,amssymb,color}

\begin{document}

\title{Correlation functions in the Schwarzian theory}

\author{Vladimir V. Belokurov}

\email{vvbelokurov@yandex.ru}

\affiliation{Lomonosov Moscow State University, Leninskie gory 1, Moscow, 119991, Russia, and Institute for Nuclear
Research of the Russian Academy of Sciences, 60th October Anniversary
Prospect 7a, Moscow, 117312, Russia}

\author{Evgeniy T. Shavgulidze}

\email{shavgulidze@bk.ru}

\affiliation{Lomonosov Moscow State University, Leninskie gory 1, Moscow, 119991, Russia}



\begin{center}
\begin{abstract}
A mathematically correct approach to study theories with infinite-dimensional groups of symmetries is presented. It is based on quasi-invariant measures on the groups. In this paper, the properties of the measure on the group of diffeomorphisms are used to evaluate the functional integrals in the Schwarzian theory.  As an important example of the application of the new technique, we  explicitly evaluate the correlation functions in the Schwarzian theory.
\end{abstract}
\end{center}
\maketitle

\centerline {\large{\textbf{CONTENTS}}}

\vspace{1cm}

\textbf{I. INTRODUCTION}................................................................................................ \textbf{3.}

\textbf{II. MATHEMATICAL PRELIMINARIES}......................................................... \textbf{4.}

\textbf{III. CORRELATION FUNCTIONS}..................................................................... \textbf{8.}

\textbf{ III.1. Mean value of $\varphi'$   }.......................................................................................... \textbf{8.}

\textbf{ III.2. Two-point correlation function   }............................................................... \textbf{10.}

\textbf{ III.3. N-point correlation functions   }.................................................................. \textbf{15.}

\textbf{IV. CONCLUDING REMARKS}......................................................................... \textbf{16.}

\textbf{ A. A quasi-invariant measure on the group of diffeomorphisms}.................... \textbf{17.}

\textbf{ B. Proof of the basic formula}............................................................................... \textbf{19.}

\textbf{ C. Properties of the function $\mathcal {E}_{\sigma}(u,\,v)$}.................................................................. \textbf{21.}

\textbf{ D. Integration in $SL(2, \textbf{R})$}...................................................................................... \textbf{24.}

\newpage

\section{ Introduction}

The Schwarzian theory is the basic element of various physical models including the SYK model and the two-dimensional dilaton gravity (see, e.g., \cite{(Kit1)}, \cite{(Kit2)}, \cite{(MS)},   \cite{(GR)}, \cite{(MNW)}, \cite{(SW)}, \cite{(KitSuh)}, \cite{(Mertens)}, and references therein).

The action of the theory is
\begin{equation}
   \label{Act2}
   I=-\frac{1}{\sigma^{2}}\int \limits _{0}^{1}\,\left[ \mathcal{S}_{\varphi}(t)+2\pi^{2}\left(\varphi'(t)\right)^{2}\right]dt\,,
\end{equation}
where
\begin{equation}
   \label{Der}
\mathcal{S}_{\varphi}(t)=
\left(\frac{\varphi''(t)}{\varphi'(t)}\right)'
-\frac{1}{2}\left(\frac{\varphi''(t)}{\varphi'(t)}\right)^2
\end{equation}
is the Schwarzian derivative.

The presence of the term $\left(\varphi'(t)\right)^{2} $ in the action (\ref{Act2}) makes the functional integral
\begin{equation}
   \label{AF}
\int \,F(\varphi)\,\exp \left\{-I \right\}\, d\varphi =\int \,F(\varphi)\,\exp\left\{\frac{1}{\sigma^{2}}\int \limits _{0}^{1}\,\left[ \mathcal{S}_{\varphi}(t)+2\pi^{2}\left(\varphi'(t)\right)^{2}\right]dt\right\}\,  d\varphi
\end{equation}
look as if it is
the integral over the  Wiener measure
\begin{equation}
   \label{Wiener}
  w_{\kappa}(d\varphi)=\exp\left\{-\frac{1}{\kappa^{2}} \int \limits _{0}^{1}\,\left(\varphi'(t)\right)^{2}dt\right\}\  d\varphi
\end{equation}
analytically continued to the point $\kappa=i \frac{\sigma}{\sqrt{2}\pi} \,.$

However, the attempt to treat (\ref{AF}) as the integral over the  Wiener measure is a misleading one.
The point is that the Wiener measure is concentrated on the trajectories that are nondifferentiable almost everywhere. The set of smooth, or even differentiable at a point, functions has zero Wiener measure.
Nevertheless, the formal representation  (\ref{Wiener}) is correct and very useful if the derivative $\varphi'(t) $ is considered in a generalized sense \cite{(Kuo)}.
However, the Schwarzian derivative cannot be understood in this way.

To evaluate correlation functions in SYK it was proposed \cite{(BAK)} to map the theory onto Liouville quantum mechanics, and to use spectral decomposition to represent functional integrals as the sum over quantum states. Although the approach turned out to be very profound and qualitatively helpful (see, also,  \cite{(BAK2)},  \cite{(MTV)}, \cite{(GR2)}), it is desirable to develop a technique of functional integration in the theory.

Mathematically correct approach is based on the quasi-invariant measure on the group of diffeomorphisms (see the next section)
\begin{equation}
   \label{Measure}
   \mu_{\sigma}(d\varphi)=\exp\left\{\frac{1}{\sigma^{2}}\int \limits _{0}^{1}\, \mathcal{S}_{\varphi}(t)\,dt  \right\}  d\varphi\,.
\end{equation}
And the functional integrals in the Schwarzian theory
should be considered as the integrals over the measure (\ref{Measure}).

In \cite{(BShExact)}, the integral for the partition function in the Schwarzian theory
\begin{equation}
   \label{PF1}
   Z_{Schw}(g)=\int \limits _{Diff^{1}([0, 1])/SL(2,\textbf{R})}\,\exp\left\{ \frac{1}{2\pi g^{2}}\int \limits _{0}^{1}\,\left[ \mathcal{S}_{\varphi}(t)+2\pi^{2}\left(\varphi'(t)\right)^{2}\right]dt\right\}\  d\varphi
\end{equation}
was explicitly evaluated with the result
\begin{equation}
   \label{Result2}
   Z_{Schw}(g)= \frac{1}{2\pi g^{3}}\,\exp\left\{ \frac{\pi}{g^{2}}\right\}\,,
\end{equation}
thereby confirming the conjecture about the exactness of the one-loop result \cite{(MS)}, \cite{(SW)}.

In this paper, we evaluate the functional integrals assigning correlation functions in the Schwarzian theory.
As the technique of functional integration over non-Wiener measures is not common knowledge, we try to make the presentation maximally explicit.

Section \textbf{II}, and  appendices A and B contain the relevant mathematical apparatus.
In Section \textbf{III},  the explicit evaluation of the correlation functions is presented. In addition, some relevant technical results are obtained in appendices C and D.  In Section \textbf{IV}, we give the concluding remarks.

\section{Mathematical preliminaries}
\label{sec:math-prelim}

For finite-dimensional groups, there is the invariant Haar measure. However, the invariant measures analogous to the Haar measure do  not exist for the infinite-dimensional groups $H.$ Nevertheless, sometimes one has succeeded in constructing the measure that is quasi-invariant with respect to the action of a more smooth subgroup $G\subset H.$ The quasi-invariance means that under the action of the subgroup $G$ the measure transforms to itself multiplied by a function $  \mathcal{R}_{g}(h)$ parametrized by the elements of the subgroup $g\in G $
$$
\mu \left(\,d(g\circ h)\,\right)=\mathcal{R}^{\mu}_{g}(h)\,\mu (\,dh\,)\,.
$$
The function $\mathcal{R}_{g}^{\mu}(h)$ is called the Radon-Nikodim derivative of the measure $\mu $
 (see, e.g., \cite{(Kuo)}, \cite{(Neretin1996)}).

The quasi-invariance of the Wiener measure (\ref{Wiener}) under the shifts of the argument of the measure by a differentiable function is the simplest example
 \cite{(Kuo)}.

The Wiener measure (\ref{Wiener}) turns out to be quasi-invariant under the group of diffeomorphisms. The proof of the quasi-invariance and the explicit form of the Radon-Nikodim derivative was first obtained in \cite{(Shepp)} (see a more simple derivation of the result, e.g., in \cite{(BSh)}).

The evaluation of the functional integrals considered in this paper is based on the  equation:
 $$
 \int \limits _{\xi(0) =\xi(1)=0}\,\exp\left\{\frac{-2\beta^{2}}{\sigma^{2}(\beta +1)}\,\frac{1}{\int \limits _{0}^{1}\,e^{\xi(t)}dt} \right\}\,w_{\sigma}(d\xi)
 $$
 \begin{equation}
   \label{Formula}
=\frac{1}{\sqrt{2\pi}\sigma}\,\exp \left\{-\frac{2}{\sigma ^{2}}\left( \log (\beta +1)\,\right)^{2} \right\}\,.
\end{equation}

It is a consequence of the quasi-invariance of the Wiener measure with respect to the action of the group of diffeomorphisms  $Diff^{3}_{+}([0,\,1])\,.$
In \cite{(BShExact)}, we postponed the proof of the equation  (\ref{Formula}) till the next paper, "Hanc marginis exiguitas non caparet." (P. Fermat).

Now, the explicit proof of the more general formula
$$
\int \limits _{\xi(0) =0,\,\xi(1)=x}\,\exp\left\{\frac{-2\beta}{\sigma^{2}}\,\left[1-\frac{e^{x}}{\beta +1} \right]\,\frac{1}{\int \limits _{0}^{1}\,e^{\xi(t)}dt }\right\}\,w_{\sigma}(d\xi)
$$
\begin{equation}
   \label{FormulaX}
  =\frac{1}{\sqrt{2\pi}\sigma}\,\exp \left\{-\frac{1}{2\sigma ^{2}}\left( x-2\log (\beta +1)\,\right)^{2} \right\}
\end{equation}
is given in \cite{(Shavgulidze2018)} ( see, also, appendix B of the present paper).

In \cite{(Shavgulidze1978)}, \cite{(Shavgulidze1988)}, \cite{(Shavgulidze2000)},  the  measures
\begin{equation}
   \label{Measure1}
   \mu_{\sigma}(X)=\int \limits _{X}\,\exp\left\{\frac{1}{\sigma^{2}}\int \limits _{0}^{1}\, \mathcal{S}_{\varphi}(t)\,dt  \right\}  d\varphi
\end{equation}
on the groups of diffeomorphisms of the interval $X\subset Diff^{1}_{+}([0,\,1])\,,$ and of the circle $X\subset Diff^{1}_{+}\left(S^{1} \right)$
were proposed.

The measures are quasi-invariant with respect to the action of the subgroups  $Diff^{3}_{+}([0,\,1])$ and $Diff^{3}_{+}(S^{1})$ respectively. The proof of the quasi-invariance and the form of the Radon-Nikodim derivatives can be obtained by some special substitution of variables \cite{(Shavgulidze1978)}, \cite{(Shavgulidze1988)}, \cite{(Shavgulidze2000)} (see, also, appendix A of the present paper).
Specifically, under the substitution
\begin{equation}
   \label{subst}
 \varphi(t)=\frac{\int \limits _{0}^{t}\,e^{\xi(\tau)}d\tau}{\int \limits _{0}^{1}\,e^{\xi(\eta)}d\eta }  \,,
\end{equation}
the measure $\mu_{\sigma}(d\varphi)$ on the group $ Diff^{1}_{+}([0,\,1])$ turns into the Wiener measure  $w_{\sigma}(d\xi)$ on $C([0,\, 1])\,.$

Consider the function $\xi\in C_{0}([0,\,1])\,.$ That is, $\xi(t)$ is a continuous function on the interval satisfying the boundary condition
$\xi(0)=0\,.$
Then
\begin{equation}
   \label{xi}
\xi(t)=\log\varphi'(t)-\log\varphi'(0)\,.
\end{equation}

The integral over the group $Diff^{1} (S^{1})$ can be transformed into the integral over the group $Diff^{1}([0,1])\,.$
Note that if we fix a point $t=0$ on the circle $S^{1}$ then it is necessary "to glue the ends of the interval". That is, to put
$\dot{\varphi}(0)=\dot{\varphi}(1)$ or
$ \xi(0)=\xi(1)=0\,.$ In this case, the function $\xi  $ is a Brownian bridge, and we denote the  corresponding functional space   by  $ C_{0,\,0}([0,\,1])\,.$ The Wiener measures on $ C_{0}([0,\,1])$ and $ C_{0,\,0}([0,\,1])$ are related by the equation
$$
w_{\sigma}(d\xi)=w^{Brown}_{\sigma}(d\xi)\frac{1}{\sqrt{2\pi}\sigma}\exp\left(-\frac{x^{2}}{2\sigma^{2}} \right)dx\,.
$$

Now, the integral  over $Diff^{1} (S^{1})$ turns into the the integral over $Diff^{1}([0,1])$ as follows:
$$
\frac{1}{\sqrt{2\pi}\sigma}\int\limits_{Diff^{1} (S^{1}) }F(\varphi)\mu_{\sigma}(d\varphi)
$$
$$
=\frac{1}{\sqrt{2\pi}\sigma} \int\limits _{C_{0,\,0}([0,1]) }F(\varphi(\xi))\, w_{\sigma}(d\xi)=\int\limits _{C_{0}([0,1]) }\delta\left(\xi(1) \right)F(\varphi(\xi)) w_{\sigma}(d\xi)
$$
\begin{equation}
   \label{Eq:Equality}
=\int\limits_{Diff^{1} ([0,1]) }\delta\left(\log\frac{\varphi'(1)}{\varphi'(0)} \right)\,F(\varphi)\,\mu_{\sigma}(d\varphi)
=\int\limits_{Diff^{1} ([0,1]) }\delta\left(\frac{\varphi'(1)}{\varphi'(0)}-1 \right)\,F(\varphi)\,\mu_{\sigma}(d\varphi)
\,.
\end{equation}

Here, we have used the equation
$$
\int\limits_{0}^{+\infty}   \delta(\log x)g(x)dx=g(1)=\int\limits_{-\infty}^{+\infty}   \delta(x-1)g(x)dx\,.
$$

     The quasi-invariance of the measure and the explicit form of the Radon-Nikodim derivative can be used to evaluate nontrivial functional integrals. For the measure $\mu $ on the interval $[0,\,1]$ in particular, the Radon-Nikodim derivative is
$$
\mathcal{R}^{\mu}_{f}(\varphi )\equiv\frac{d\mu_{\sigma}^{f}}{d\mu_{\sigma}}(\varphi)=\frac{1}{\sqrt{f'(0)f'(1)}}
$$
\begin{equation}
   \label{RadNik}
  \times \exp\left\{ \frac{1}{\sigma^{2}}\left[   \frac{f''(0)}{f'(0)}\varphi'(0)-  \frac{f''(1)}{f'(1)}\varphi'(1)\right]   +        \frac{1}{\sigma^{2}}\int \limits _{0}^{1}\, \mathcal{S}_{f}\left(\varphi(t)\right)\,\left(\varphi' (t)\right)^{2}\,dt  \right\} \,,
\end{equation}
where
$$
\mu_{\sigma}^{f}(X)=\mu_{\sigma}(f\circ X)\,.
$$
Here,  the well known property of the Schwarzian derivative :
$$
\mathcal{S}_{f\circ \varphi}(t)= \mathcal{S}_{f}(\varphi(t))\left(\varphi' (t)\right)^{2}+ \mathcal{S}_{\varphi}(t)\,,\ \ \ \ (\,f\circ \varphi\,)(t)=f\left( \varphi(t)\right)\,.
$$
has been used.

Thus, for functional integrals over the measure  $\mu \,, $ we have
  $$
  \int \limits _{Diff^{1}([0, 1])}\,F(\varphi)\mu_{\sigma}(d\varphi)=\frac{1}{\sqrt{f'(0)f'(1)}}\int \limits _{Diff^{1}([0, 1])}\,F(f\circ\varphi)
  $$
  \begin{equation}
   \label{FI1}
  \times \exp\left\{ \frac{1}{\sigma^{2}}\left[   \frac{f''(0)}{f'(0)}\varphi'(0)-  \frac{f''(1)}{f'(1)}\varphi'(1)\right]   +        \frac{1}{\sigma^{2}}\int \limits _{0}^{1}\, \mathcal{S}_{f}\left(\varphi(t)\right)\,\left(\varphi' (t)\right)^{2}\,dt  \right\} \,\mu_{\sigma}(d\varphi)\,.
\end{equation}

In what follows, we assume the function $f$ to be
\begin{equation}
   \label{f}
f(t)=f_{\alpha}(t)=\frac{1}{2}\left[ \frac{1}{\tan\frac{\alpha}{2}}\tan\left(\alpha(t-\frac{1}{2}) \right)+1 \right]\,.
\end{equation}
In this case,
\begin{equation}
   \label{tan}
f'_{\alpha}(0)=f'_{\alpha}(1)=\frac{\alpha}{\sin \alpha}\,,
\ \ \ -\frac{f''_{\alpha}(0)}{f'_{\alpha}(0)}=\frac{f''_{\alpha}(1)}{f'_{\alpha}(1)}=2\alpha\tan\frac{\alpha}{2}\,, \ \ \ \mathcal{S}_{f_{\alpha}}(t)=2\alpha^{2}\,,
\end{equation}
and the equation (\ref{FI1}) looks like:
$$
\frac{\alpha}{\sin \alpha}\,\int \limits _{Diff^{1}([0, 1])}F(\varphi)\mu_{\sigma}(d\varphi) =\int \limits _{Diff^{1}([0, 1])}F(f_{\alpha}(\varphi))
$$
\begin{equation}
   \label{FI2}
\times\,\exp\left\{-\frac{2\alpha}{\sigma^{2}}\tan\frac{\alpha}{2}\left(\varphi'(0)+\varphi'(1)\right)\right\}
 \exp\left\{\frac{2\alpha^{2}}{\sigma^{2}}\int \limits _{0}^{1}\left(\varphi' (t)\right)^{2}dt  \right\} \mu_{\sigma} (d\varphi)\,.
\end{equation}

Generally speaking, the functional integrals (\ref{FI2})  converge for $ 0\leq \alpha < \pi \,, $ and diverge for $ \alpha = \pi \,. $

The Schwarzian action is invariant under the noncompact group $SL(2, \textbf{R})$. Therefore, integrating over the quotient space $Diff^{1}([0, 1])/SL(2,\textbf{R}) $
we exclude the infinite volume of the group $SL(2, \textbf{R})$ and get the finite results for functional integrals in the Schwarzian theory.
In our approach, we evaluate regularized $(\alpha<\pi)$ functional integrals over the group  $Diff^{1}([0, 1])$ and then normalize them to the corresponding integrals over the group $SL(2, \textbf{R})\,.$

In particular, in \cite{(BShExact)}, to get the partition function (\ref{Result2}), we first evaluated the regularized integral
$$
Z_{\alpha}(\sigma)=\int \limits _{  Diff^{1}(S^{1}) }
\exp\left\{-I\right\}\,\exp \left\{\frac{-2\left[ \pi^{2}-\alpha^{2}\right]}{\sigma^{2}}\int \limits _{0}^{1}\left(\varphi' (t)\right)^{2}\,dt  \right\}  d\varphi
$$
\begin{equation}
   \label{RegPF}
=\int \limits _{ Diff^{1}(S^{1})}\exp\left\{\frac{2\alpha^{2}}{\sigma^{2}}\int \limits _{0}^{1}\left(\varphi' (t)\right)^{2}\,dt  \right\} \mu_{\sigma} (d\varphi)\,,
\end{equation}
and then divided it by the regularized volume of the group $SL(2,\textbf{R})$
\begin{equation}
   \label{V}
V_{\alpha}(\sigma)=\int \limits _{SL(2,\textbf{R})}\exp \left\{\frac{-2\left[ \pi^{2}-\alpha^{2}\right]}{\sigma^{2}}\int \limits _{0}^{1}\left(\varphi' (t)\right)^{2}\,dt  \right\}  d\mu _{H}\,.
\end{equation}
Note that the functional measure in the equation (\ref{RegPF}) and the Haar measure $d\mu _{H}$ on the group $SL(2, \textbf{R})$  in the equation (\ref{V}) are regularized in the same manner.

For the Schwarzian partition function, we take the limit
\begin{equation}
   \label{Ratio2}
  Z(\sigma) = \lim \limits_{\alpha\rightarrow \pi - 0}\ \frac{Z_{\alpha}(\sigma)}{V_{\alpha}(\sigma)} \,.
\end{equation}

In the next section, the quasi-invariance of the measure (\ref{Measure}) is used to evaluate the functional integrals assigning the correlation functions in the Schwarzian theory.

\section{ Correlation functions}
\label{sec:two-point}

\textbf{III.1. Mean value of $\varphi'$ }

\vspace{0.5cm}

First we recall the main steps of the evaluation of the partition function in the Schwarzian theory \cite{(BShExact)}.

If we take the function
 $F$  in the equation  (\ref{FI2})     to be
\begin{equation}
   \label{F}
F(f(\varphi))=F_{1}(f(\varphi))=\exp\left\{\frac{4\alpha}{\sigma^{2}}\tan\frac{\alpha}{2}\,\varphi'(0)\right\}\,,
\end{equation}
and note that
$$
\varphi'(0)=\frac{1}{f'(0)}u'(0)=\frac{\sin \alpha}{\alpha}u'(0)\,,
$$
for $u(t)=f(\varphi(t))$, then
\begin{equation}
   \label{F1(u)}
F_{1}(u)=\exp\left\{\frac{8\,\sin ^{2}\frac{\alpha}{2}}{\sigma^{2}}\,u'(0)\right\}\,.
\end{equation}
Now  the regularized partition function has the form
\begin{equation}
   \label{RegPF1}
Z_{\alpha}(\sigma)=\frac{\alpha}{\sin \alpha}\,\int \limits _{\dot{\varphi}(0) =\dot{\varphi}(1)}\,\exp\left\{\frac{8\,\sin ^{2}\frac{\alpha}{2}}{\sigma^{2}}\,\dot{\varphi}(0)\right\}\,\mu_{\sigma}(d\varphi)\,.
\end{equation}
Under the substitution (\ref{subst})
it turns into
\begin{equation}
   \label{RegPF2}
 Z_{\alpha}(\sigma)=\frac{\alpha}{\sin \alpha}\int \limits _{\xi(0) =\xi(1)=0}\,\exp\left\{\frac{8\,\sin ^{2}\frac{\alpha}{2}}{\sigma^{2}}\,\frac{1}{\int \limits _{0}^{1}\,e^{\xi(\tau)}d\tau } \right\}\,w_{\sigma}(d\xi)\,.
\end{equation}

To evaluate the functional integral explicitly we use the equation (\ref{Formula}).  Instead of $\beta \,,$ we should substitute a solution of the equation
\begin{equation}
   \label{BetaA}
  \frac{2\beta^{2}}{\sigma^{2}( \beta +1)}= -\frac{8\sin^{2}\frac{\alpha}{2}}{\sigma ^{2}}\,,
\end{equation}
We take the following one:
\begin{equation}
   \label{BetaAlfa}
(\beta +1)=e^{i\alpha}\,.
\end{equation}
As the result, we obtain
\begin{equation}
   \label{RegPF3}
 Z_{\alpha}(\sigma)=\frac{\alpha}{\sin \alpha}\,\frac{1}{\sqrt{2\pi} \sigma}\,\exp\left\{ \frac{2\alpha^{2}}{\sigma^{2}}\right\}\,,
\end{equation}
with the asymptotic form at $\alpha\rightarrow\pi $
\begin{equation}
   \label{ZAs}
   Z^{As}_{\alpha}(\sigma)= \frac{\pi }{\pi-\alpha}\,\frac{1}{\sqrt{2\pi} \sigma}\,\exp\left\{ \frac{2\pi^{2}}{\sigma^{2}}\right\}\,.
\end{equation}

The asymptotic form of the $\alpha -$regularized volume of the group $ SL(2,\textbf{R}) $ (see appendix D)
looks like
\begin{equation}
   \label{VAs}
V^{As}_{\alpha; SL(2,\textbf{R})}(\sigma)= \frac{ \sigma^{2}}{2\left[\pi-\alpha\right]}\,.
\end{equation}

According to the equation (\ref{Ratio2}), the Schwarzian partition function has the form
\begin{equation}
   \label{Result2}
   Z(\sigma)= \frac{\sqrt{2\pi}}{\sigma^{3}}\,\exp\left\{ \frac{2\pi ^{2}}{\sigma^{2}}\right\}\,.
\end{equation}

Consider now the $\alpha-$regularized mean value of $\varphi'$
$$
\Phi^{\alpha}=\frac{1}{\sqrt{2\pi}\sigma}\int\limits_{Diff^{1} (S^{1}) }\,\varphi'(0)
\exp\left\{\frac{2\alpha^{2}}{\sigma^{2}}\int \limits _{0}^{1}\left(\varphi' (\tau)\right)^{2}\,d\tau  \right\} \mu_{\sigma} (d\varphi)\,.
$$

Note that it is $\varphi'$, but not $\varphi$, that is the dynamical variable in the theory given by the action (\ref{Act2}).

After the substitution (\ref{subst}),
it is written as
\begin{equation}
   \label{RegFi}
\Phi^{\alpha}=\int \limits _{\xi(0) =\xi(1)=0}\,\frac{1}{\int \limits _{0}^{1}e^{\xi(\tau)}d\tau }\,\exp\left\{\frac{8\,\sin ^{2}\frac{\alpha}{2}}{\sigma^{2}}\,\frac{1}{\int \limits _{0}^{1}\,e^{\xi(\tau)}d\tau }\right\}\,w_{\sigma}(d\xi)\,.
\end{equation}
Having in mind the equations (\ref{Formula}), (\ref{BetaA}), and (\ref{BetaAlfa}), we get
\begin{equation}
   \label{RegFi2}
\Phi^{\alpha}=\frac{4}{\sqrt{2\pi}\sigma^{3}}\,\frac{\ln(\beta +1)}{(\beta +1)}\,\exp\left\{-\frac{2\ln^{2}(\beta +1)}{\sigma^{2}} \right\}=
\frac{\alpha}{\sin \alpha}\,\frac{1}{\sqrt{2\pi} \sigma}\,\exp\left\{ \frac{2\alpha^{2}}{\sigma^{2}}\right\}\,.
\end{equation}
Asymptotically, it looks like
\begin{equation}
   \label{FiAs}
   \Phi^{As}= \frac{\pi }{\pi-\alpha}\,\frac{1}{\sqrt{2\pi} \sigma}\,\exp\left\{ \frac{2\pi^{2}}{\sigma^{2}}\right\}\,.
\end{equation}

The asymptotic form of the $\Phi^{\alpha} $ on the group $ SL(2,\textbf{R}) $ (see  appendix D) is
\begin{equation}
   \label{VAs}
\Phi^{As}_{ SL(2,\textbf{R})}= \frac{ \sigma^{2}}{2\left[\pi-\alpha\right]}\,.
\end{equation}

Now the normalized mean value of $\varphi '$ has the form
\begin{equation}
   \label{Result2}
   \Phi = \lim \limits_{\alpha\rightarrow \pi - 0}\ \frac{\Phi^{As}}{\Phi^{As}_{ SL(2,\textbf{R})}}= \frac{\sqrt{2\pi}}{\sigma^{3}}\,\exp\left\{ \frac{2\pi ^{2}}{\sigma^{2}}\right\}\,.
\end{equation}

\vspace{0.5cm}

\textbf{III.2. Two-point correlation function}

\vspace{0.5cm}

Define the $\alpha -$regularized two-point correlation function as
$$
G^{\alpha}_{2}\left(0,\,t \right)=\frac{1}{\sqrt{2\pi}\sigma}\int\limits_{Diff^{1} (S^{1}) }\varphi'(t)\,\varphi'(0)
\exp\left\{\frac{2\alpha^{2}}{\sigma^{2}}\int \limits _{0}^{1}\left(\varphi'(\tau) \right)^{2}\,d\tau  \right\} \mu_{\sigma} (d\varphi)
$$
\begin{equation}
   \label{Def}
=\int\limits_{Diff^{1} ([0,\,1]) }\varphi'(t)\,\varphi'(0) \, \delta\left(\frac{\varphi'(1)}{\varphi'(0)}-1 \right)\exp\left\{\frac{2\alpha^{2}}{\sigma^{2}}\int \limits _{0}^{1}\left(\varphi'(\tau) \right)^{2} d\tau  \right\} \mu_{\sigma} (d\varphi)\,.
\end{equation}

 By the special choice of the function $F$ in (\ref{FI2}), we identify the integrands in (\ref{Def}) and in the right-hand side of the equation (\ref{FI2}).

Represent the function $F$ in the form
$$
F(\varphi )=F_{4}(\varphi )\, F_{3}(\varphi )\,F_{2}(\varphi )\, F_{1}(\varphi)\,,
$$
where
$$
F_{1}(f\circ \varphi)=\exp\left\{\frac{4\alpha}{\sigma^{2}}\tan\frac{\alpha}{2}\,\varphi'(0)\right\}\,,\ \ \ F_{2}(f\circ \varphi)=\delta\left(\varphi'(1)-\varphi'(0) \right)\,,
$$
$$ F_{3}(f\circ \varphi)=\left(\varphi'(0) \right)^{2} \,,\ \ \  F_{4}(f\circ \varphi)=\varphi'(t)\,.
$$

To use the equation (\ref{FI2}), it is necessary  to find $F_{i}(\varphi)\,.$
$F_{1}(\varphi)$ was found above to be
\begin{equation}
   \label{F1}
F_{1}(\varphi)=\exp\left\{\frac{8\,\sin ^{2}\frac{\alpha}{2}}{\sigma^{2}}\,\varphi'(0)\right\}\,.
\end{equation}
To find $F_{2}(\varphi)$ and $F_{3}(\varphi)\,,$ note that for $\chi(t)=f(\varphi(t))\,, $
$$
\chi'(0)=\frac{\alpha}{\sin \alpha}\,\varphi'(0)\,,\ \ \ \chi'(1)=\frac{\alpha}{\sin \alpha}\,\varphi'(1)\,,
$$
so
\begin{equation}
   \label{F2}
F_{2}(\varphi)=\frac{\alpha}{\sin \alpha}\,\delta\left(\varphi'(1)-\varphi'(0) \right)\,,
\end{equation}
and
\begin{equation}
   \label{F3}
F_{3}(\varphi)=\frac{\sin^{2} \alpha}{\alpha^{2}}\,\left(\varphi'(0) \right)^{2} \,.
\end{equation}
 $F_{4}(\varphi)$ looks more complicated:
$$
F_{4}\left(\chi(t)\right)=F_{4}\left(f( \varphi(t))\right)=\varphi'(t)=\frac{1}{f'\left(f^{-1}\left(\chi(t) \right) \right)}\,\chi'(t)=\left(f^{-1}\right)'\left(\chi(t) \right)\,\chi'(t)\,.
$$
For the function $y=f_{\alpha}(x) $ given by the equation (\ref{f}),
$$
x=f_{\alpha}^{-1}(y)=\frac{1}{\alpha}\arctan\left[\tan \frac{\alpha}{2}\,(2y-1)\right]+\frac{1}{2}\,.
$$
And finally,
\begin{equation}
   \label{F4}
F_{4}(\varphi)= \frac{\frac{2}{\alpha}\tan \frac{\alpha}{2}}{1+\tan^{2} \frac{\alpha}{2}\,\left(2\varphi(t)-1 \right)^{2}}\,\,\varphi'(t)\,.
\end{equation}

Thus for the correlation function, we have
$$
G^{\alpha}_{2}\left(0,\,t \right)=\int\limits_{Diff^{1} ([0,\,1]) }\frac{\frac{2}{\alpha}\tan \frac{\alpha}{2}}{1+\tan^{2} \frac{\alpha}{2}\,\left(2\varphi(t)-1 \right)^{2}}
$$
$$
\times\varphi'(t)\,\varphi'(0) \, \delta\left(\frac{\varphi'(1)}{\varphi'(0)}-1 \right)\,\exp\left\{\frac{8\,\sin ^{2}\frac{\alpha}{2}}{\sigma^{2}}\,\varphi'(0)\right\}\,\mu_{\sigma}(d\varphi)
$$
$$
=\int\limits_{0}^{1}\,\frac{\frac{2}{\alpha}\tan \frac{\alpha}{2}}{1+\tan^{2} \frac{\alpha}{2}\,\left(2x-1 \right)^{2}}\,dx\,\int\limits_{Diff^{1} ([0,\,1]) }\,\delta \left(x-\varphi (t) \right)
$$
\begin{equation}
   \label{2pGF}
\times\varphi'(t)\,\varphi'(0) \, \delta\left(\frac{\varphi'(1)}{\varphi'(0)}-1 \right)\,\exp\left\{\frac{8\,\sin ^{2}\frac{\alpha}{2}}{\sigma^{2}}\,\varphi'(0)\right\}\,\mu_{\sigma}(d\varphi)\,.
\end{equation}

The substitution (\ref{subst}) in the above equation gives the correlation function in terms of the Wiener integral:
$$
G^{\alpha}_{2}\left(0,\,t \right)=\int\limits_{0}^{1}\frac{\frac{2}{\alpha}\tan \frac{\alpha}{2}}{1+\tan^{2} \frac{\alpha}{2}\,\left(2x-1 \right)^{2}}dx\int\limits_{C_{0} ([0,\,1]) }\delta \left(x\int \limits _{0}^{1}e^{\xi(\tau)}d\tau  - \int \limits _{0}^{t}e^{\xi(\tau)}d\tau  \right)
$$
\begin{equation}
   \label{2pGFasWI}
\times \,\frac{e^{\xi(t)}}{\int \limits _{0}^{1}e^{\xi(\tau)}d\tau }\ \delta\left(\,e^{\xi(1)} - 1\, \right)\,\exp\left\{\frac{8\,\sin ^{2}\frac{\alpha}{2}}{\sigma^{2}}\,\frac{1}{\int \limits _{0}^{1}e^{\xi(\tau)}d\tau } \right\}\,w_{\sigma}(d\xi)\,.
\end{equation}

Divide the interval $[0,\,1]$ into the two intervals $[0,\,t]$ and $[t,\,1]\,.$ The substitution
\begin{equation}
   \label{xieta}
\xi(\tau)=\eta_{1}\left(\frac{\tau}{t} \right)\,, \ \ \tau\leq t\,; \ \ \ \ \ \ \ \ \ \  \xi(\tau)=\eta_{1}(1) +\eta_{2}\left(\frac{\tau-t}{1-t} \right)\,,\ \ \tau>t
\end{equation}
transforms  the Wiener integral over the measure $w_{\sigma}(d\xi)$ into the two Wiener integrals over the measures
$$
w_{\sigma \sqrt{t}}(d\eta_{1})\,w_{\sigma \sqrt{1-t}}(d\eta_{2})\,.
$$
To verify this statement, note that
$$
\int \limits _{0}^{t}\left(\xi'(\tau) \right)^{2}\,d\tau=\frac{1}{t}\int \limits _{0}^{1}\left(\eta'_{1}(\tau) \right)^{2}\,d\tau\,, \ \ \ \ \ \ \ \ \ \ \int \limits _{t}^{1}\left(\xi'(\tau) \right)^{2}\,d\tau=\frac{1}{1-t}\int \limits _{0}^{1}\left(\eta'_{2}(\tau) \right)^{2}\,d\tau\,.
$$

To return to the integrals over the group of diffeomorphisms, consider the functions
\begin{equation}
   \label{psi}
 \psi_{1}(t)=\frac{\int \limits _{0}^{t}e^{\eta_{1}(\tau)}d\tau }{\int \limits _{0}^{1}e^{\eta_{1}(\tau)}d\tau }\,, \ \ \ \ \ \ \ \ \ \ \psi_{2}(t)=\frac{\int \limits _{0}^{t}e^{\eta_{2}(\tau)}d\tau }{\int \limits _{0}^{1}e^{\eta_{2}(\tau)}d\tau }\,.
\end{equation}

The useful relations can be obtained directly from the above definitions
$$
\varphi(t)=\frac{t\psi'_{2}(0)}{t\psi'_{2}(0)+(1-t)\psi'_{1}(1)}\,,
\ \ \ \ \ \ \
\varphi' (t)=\frac{\psi'_{1}(1)\psi'_{2}(0)}{t\psi'_{2}(0)+(1-t)\psi'_{1}(1)}\,,
$$
$$
\varphi' (0)=\frac{\psi'_{1}(0)\psi'_{2}(0)}{t\psi'_{2}(0)+(1-t)\psi'_{1}(1)}\,,
\ \ \ \ \ \ \
\varphi' (1)=\frac{\psi'_{1}(1)\psi'_{2}(1)}{t\psi'_{2}(0)+(1-t)\psi'_{1}(1)}\,.
$$

Therefore, the two-point correlation function has the form of the double functional integral
$$
G^{\alpha}_{2}\left(0,\,t \right)= \frac{1}{t\left(1-t\right)}\int\limits_{0}^{1}\,\frac{\frac{2}{\alpha}\tan \frac{\alpha}{2}}{1+\tan^{2} \frac{\alpha}{2}\,\left(2x-1 \right)^{2}}\,dx\int\limits_{Diff^{1} ([0,\,1]) }\int\limits_{Diff^{1} ([0,\,1]) }\left(\psi'_{1}(1)\right)^{2}\left(\psi'_{2}(1)\right)^{2}
$$
$$
\times\,\delta \left(\psi'_{2}(0)-\frac{(1-t)}{t}\frac{x}{1-x}\psi'_{1}(1)\right)
\,\delta \left(\psi'_{1}(0)- \frac{t}{1-t}\frac{1-x}{x}\psi'_{2}(1) \right)
$$
\begin{equation}
   \label{2pGFasWImu}
\times \,\exp\left\{\frac{8\,\sin ^{2}\frac{\alpha}{2}}{\sigma^{2}}\,\frac{(1-x)}{( 1-t) }\psi'_{2}(1) \right\}\,\mu_{\sigma \sqrt{t}}(d\psi_{1})\,\mu_{\sigma \sqrt{1-t}}(d\psi_{2})\,.
\end{equation}

Define the function $\mathcal {E}_{\sigma}(u,\,v) $ by the equation
\begin{equation}
   \label{E}
\mathcal {E}_{\sigma}(u,\,v)=\int\limits_{Diff^{1} ([0,1]) }\,\delta\left(\varphi'(0)-u \right)\,\delta\left(\varphi'(1)-v \right)\,\mu_{\sigma}(d\varphi)\,.
\end{equation}

We can rewrite the equation (\ref{2pGFasWImu}) as
$$
G^{\alpha}_{2}\left(0,\,t \right)= \frac{1}{t\left(1-t\right)}\int\limits_{0}^{1}\,\frac{\frac{2}{\alpha}\tan \frac{\alpha}{2}}{1+\tan^{2} \frac{\alpha}{2}\,\left(2x-1 \right)^{2}}dx\int\limits_{0}^{+\infty}\int\limits_{0}^{+\infty}v_{1}^{2}\,v_{2}^{2}\exp\left\{\frac{8\,\sin ^{2}\frac{\alpha}{2}}{\sigma^{2}}\,\frac{(1-x)}{( 1-t) }v_{2} \right\}
$$
\begin{equation}
   \label{GFasE}
\times\,\mathcal {E}_{\sigma \sqrt{t}}\left(\frac{t}{1-t}\frac{1-x}{x}v_{2},\,v_{1}\right)\,\mathcal {E}_{\sigma \sqrt{1-t}}\left(\frac{1-t}{t }\frac{x}{1-x}v_{1},\,v_{2}\right)\,dv_{1}\,dv_{2}\,.
\end{equation}

In terms of variables
$$
V_{1}=\frac{x}{t}\,v_{1}\,, \ \ \ \ \ V_{2}=\frac{1-x}{1-t}\,v_{2}\,,
$$
the correlation function looks like
$$
G_{2}^{\alpha}\left(0,\,t \right)= t^{2}(1-t)^{2}\int\limits_{0}^{1}\,\frac{1}{x^{3}(1-x)^{3}}\,\frac{\frac{2}{\alpha}\tan \frac{\alpha}{2}}{1+\tan^{2} \frac{\alpha}{2}\,\left(2x-1 \right)^{2}}\,dx\,\int\limits_{0}^{+\infty}\int\limits_{0}^{+\infty}V_{1}^{2}\,V_{2}^{2}
$$
\begin{equation}
   \label{GFV}
\times\,\exp\left\{\frac{8\sin^{2}\frac{\alpha}{2}}{\sigma^{2}}
V_{2} \right\}\mathcal {E}_{\sigma \sqrt{t}}\left(\frac{t}{x}V_{2},\,\frac{t}{x}V_{1}\right)\,\mathcal {E}_{\sigma \sqrt{1-t}}\left( \frac{1-t}{1-x} V_{1},\,\frac{1-t}{1-x}V_{2}\right)\,dV_{1}\,dV_{2}\,.
\end{equation}

In appendix C, we study the properties of the function $\mathcal {E} $ and, in particular, obtain the equations
\begin{equation}
   \label{Esymm}
\mathcal {E}_{\sigma }\left(u,\,v\right)=\mathcal {E}_{\sigma }\left(\sqrt{uv},\,\sqrt{uv}\right)\,\exp \left\{ -\frac{2}{\sigma^{2}}\left(\sqrt{u}-\sqrt{v}\right)^{2} \right\}\,,
\end{equation}
$$
\mathcal {E}_{\sigma }\left(u,\,u\right)=\frac{2\sqrt{2}}{u\,\pi^{\frac{3}{2}}\sigma^{3}}\,\exp\left\{\frac{2\pi^{2}-4\,u}{\sigma^{2}} \right\}
$$
\begin{equation}
   \label{Eint}
\times\int\limits_{0}^{+\infty}\,\exp\left\{-\frac{2}{\sigma^{2}}\left(2\,u\,\cosh\tau+\tau^{2}\right) \right\}\,\sin\left(\frac{4\pi\tau}{\sigma^{2}} \right)\,\sinh(\tau)\,d\tau \,.
\end{equation}

Now, using the equations (\ref{Esymm}), (\ref{Eint}) and the substitution $ z=(2x-1)\,\tan \frac{\alpha}{2}\,,$    we get the following representation for the correlation function in terms of the ordinary integrals:
$$
G_{2}^{\alpha}\left(0,\,t \right)= \frac{8^{3}}{\pi^{3}\sigma^{6}}\,\frac{1}{\sqrt{t(1-t)}}\,\exp\left\{ \frac{2\pi^{2}}{\sigma^{2}\,t(1-t)}\right\}\,\int\limits_{0}^{+\infty}\,\int\limits_{0}^{+\infty}\,H(\alpha;\,\tau,\theta)
$$
\begin{equation}
   \label{GH}
 \times\,\exp\left\{-\frac{2}{\sigma^{2}}\left[\frac{\tau^{2}}{t}+\frac{\theta^{2}}{(1-t)}\right]\right\}\sin\left(\frac{4\pi\,\tau}{\sigma^{2}\,t}\right)
\sin\left(\frac{4\pi\,\theta}{\sigma^{2}\,(1-t)}\right)\sinh(\tau)\sinh(\theta)\,d\tau\,d\theta\,,
\end{equation}
where
$$
H(\alpha;\,\tau,\theta)=\frac{1}{\alpha}\,\int\limits_{-\tan\frac{\alpha}{2}}^{+\tan\frac{\alpha}{2}}\,\frac{1}{\left[1-z^{2}\cot^{2}\frac{\alpha}{2}\right]^{2}}\,
\frac{dz}{1+z^{2}}
$$
$$
\times \,\int\limits_{0}^{+\infty}\int\limits_{0}^{+\infty}\,dV_{1}\,dV_{2}\,\,V_{1}\,V_{2}
\,\exp
\left\{-\frac{8}{\sigma^{2}}\frac{1}{\left(1-z^{2}\cot^{2}\frac{\alpha}{2}\right)}\left((1+z^{2})V_{2}\cos^{2}\frac{\alpha}{2}+V_{1}\right) \right\}
$$
\begin{equation}
   \label{H}
\times\,\exp
\left\{-\frac{8}{\sigma^{2}}\left(\frac{\cosh(\tau)}{1+z\cot\frac{\alpha}{2}}+\frac{\cosh(\theta)}{1-z\cot\frac{\alpha}{2}} \right) \sqrt{V_{1}V_{2}}\right\}\,.
\end{equation}

The substitutions
$$
V_{1}=\rho \sin^{2}\omega\,, V_{2}=\rho \cos^{2}\omega\,,
$$
and $y=\tan \omega $ reduce the equation (\ref{H}) to the table integrals (see, e.g., \cite{(Prud)} n. 2.2.9.11 ) with the result
$$
H(\alpha;\,\tau,\theta)=12\left(\frac{\sigma^{2}}{8}\right)^{4}\frac{1}{\left(\cosh (\tau) +\cosh (\theta)\right)^{4}}
$$
\begin{equation}
   \label{Has}
\times\left\{-\log\sin\alpha +\log (\cosh \tau +\cosh \theta)-\frac{3}{2}\log 2+\log \sigma -\frac{11}{3}+O(\pi-\alpha)\right\}\,.
\end{equation}

 The asymptotics of the two-point correlation function at $\alpha\rightarrow \pi $ has the form
$$
G_{2}^{As}\left(0,\,t \right)= \frac{3\sigma^{2}}{2\pi^{3}}\,\frac{\left( -\log\sin\alpha\right)}{\sqrt{t(1-t)}}\,\exp\left\{ \frac{2\pi^{2}}{\sigma^{2}\,t(1-t)}\right\}\int\limits_{0}^{+\infty}\int\limits_{0}^{+\infty}\exp\left\{-\frac{2}
{\sigma^{2}}\left[\frac{\tau^{2}}{t}+\frac{\theta^{2}}{(1-t)}\right]\right\}
$$
\begin{equation}
   \label{Gas}
 \times
 \sin\left(\frac{4\pi\tau}{\sigma^{2}t}\right)
\sin\left(\frac{4\pi\theta}{\sigma^{2}(1-t)}\right)\frac{\sinh(\tau)\sinh(\theta)}{\left(\cosh (\tau) +\cosh (\theta)\right)^{4}}d\tau\,d\theta\,.
\end{equation}

Define the normalized two-point correlation function $G _{2}(0, t) $  as the limit
\begin{equation}
   \label{G2lim}
G_{2}\left(0, t\right)=\lim \limits_{\alpha\rightarrow\pi - 0}\  \frac{ G_{2}^{\alpha}\left(0,\,t \right)}{ G^{\alpha} _{2;\,SL(2, \textbf{R})}\left(0,\,\frac{1}{2}\right)}\,.
\end{equation}
Here, the correlation function on the group $SL(2, \textbf{R}) $ at the symmetrical points is chosen as the normalizing factor. It is evaluated in  appendix D with the result
$$
G^{As} _{2;\,SL(2, \textbf{R})}\left(0,\,\frac{1}{2}\right)=-\pi\,\log(\pi-\alpha)\,.
$$

Thus we have
$$
G_{2}\left(0,\,t \right)= \frac{3\sigma^{2}}{2\pi^{4}}\,\frac{1}{\sqrt{t(1-t)}}\,\exp\left\{ \frac{2\pi^{2}}{\sigma^{2}t(1-t)}\right\}
\int\limits_{0}^{+\infty}\,\int\limits_{0}^{+\infty}\,\exp\left\{-\frac{2}{\sigma^{2}}\left[\frac{\tau^{2}}{t}+\frac{\theta^{2}}{(1-t)}\right]\right\}
$$
\begin{equation}
   \label{GFfinal}
 \times\,
 \sin\left(\frac{4\pi\,\tau}{\sigma^{2}\,t}\right)
\sin\left(\frac{4\pi\,\theta}{\sigma^{2}\,(1-t)}\right)\frac{\sinh(\tau)\sinh(\theta)}{{\left(\cosh (\tau) +\cosh (\theta)\right)^{4}} }\,d\tau\,d\theta\,.
\end{equation}

From the above equation, it follows that the correlation function is singular at $t\rightarrow 0$, and $t\rightarrow 1\,.$ Its form is presented at Fig.1.

\begin{figure}
\includegraphics [height=0.3\textwidth]{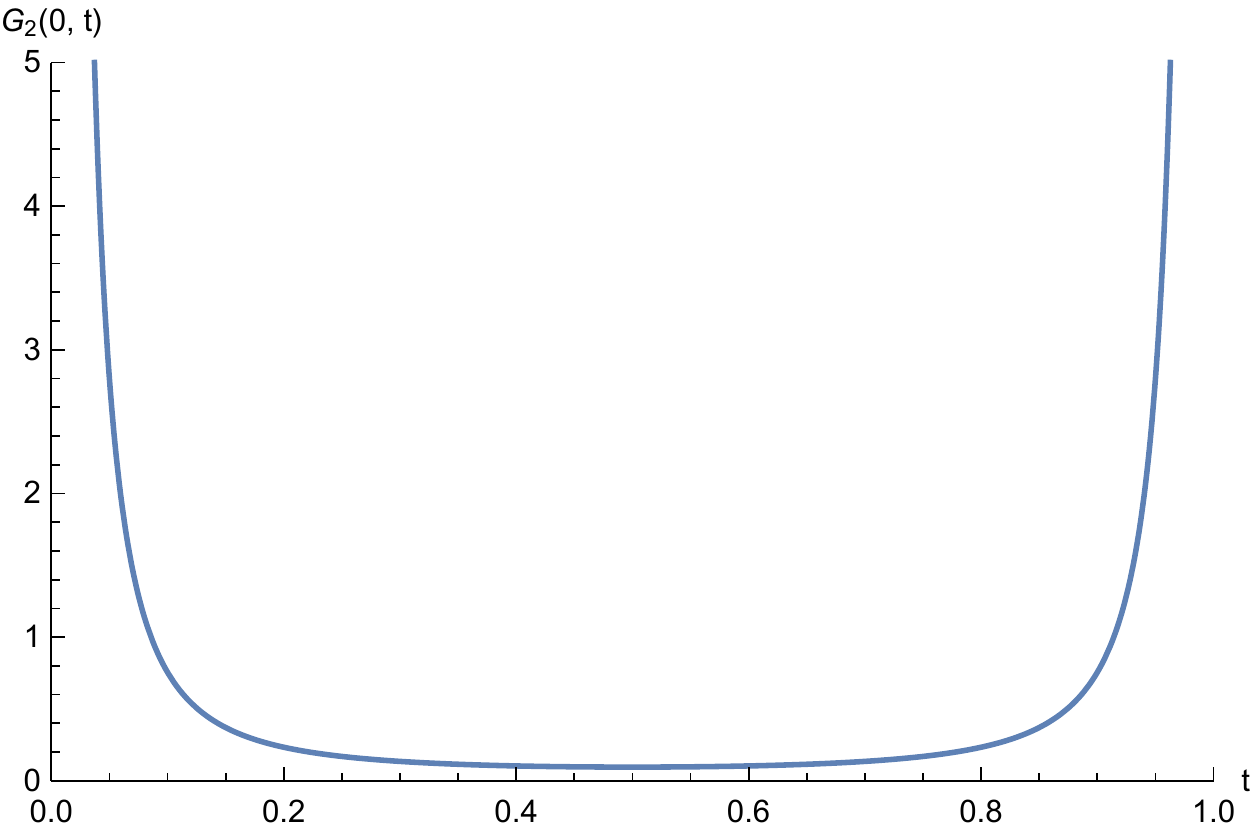}
\caption{The form of the correlation function  $G_{2}\left(0,\,t \right) $ for $\sigma=2\pi\,.$}
\end{figure}

\vspace{0.5cm}

\textbf{III.3. N-point correlation functions}

\vspace{0.5cm}

The method described in detail in the above subsection can be used to evaluate the N-point correlation function given by the functional integral
 $$
G^{\alpha}_{N}\left(0,\,t_{1},...t_{N-1} \right)
$$
\begin{equation}
   \label{DefN}
=\int\limits_{Diff^{1} ([0,\,1]) }\varphi'(0) \, \varphi'(t_{1})\,...\varphi'(t_{N-1})\,\delta\left(\frac{\varphi'(1)}{\varphi'(0)}-1 \right)\exp\left\{\frac{2\alpha^{2}}{\sigma^{2}}\int \limits _{0}^{1}\left(\varphi'(\tau) \right)^{2} d\tau  \right\} \mu_{\sigma} (d\varphi)\,.
\end{equation}

Let
$$
t_{0}=0<t_{1}<...<t_{N-1}< t_{N}=1;\ \ \ \ \ \ \ \ x_{0}=0,\ x_{N}=1\,,
$$
and denote
$$
\Delta t_{n}=t_{n}-t_{n-1}\,,\ \ \ \ \Delta x_{n}=x_{n}-x_{n-1}\,;\ \ \ \ \chi_{\alpha}(x_{n})=\frac{\frac{2}{\alpha}\tan \frac{\alpha}{2}}{1+\tan^{2} \frac{\alpha}{2}\,\left(2x_{n}-1 \right)^{2}}\,, \ \ n=1,...N\,.
$$
Then the N-point correlation function is written as
 $$
 G^{\alpha}_{N}\left(0,\,t_{1},...t_{N-1} \right)=\left(\Delta t_{1}\cdot\cdot\cdot\Delta t_{N} \right)^{2}
 $$
 $$
\times\,\int\limits_{0}^{1}\chi_{\alpha}(x_{N-1})dx_{N-1} \int\limits_{0}^{x_{N-1}}\chi_{\alpha}(x_{N-2})dx_{N-2}\cdot\cdot\cdot\int\limits_{0}^{x_{2}}\chi_{\alpha}(x_{1})dx_{1}\frac{1}{\left(\Delta x_{1}\cdot\cdot\cdot\Delta x_{N} \right)^{3}}
$$
$$
\times\int\limits_{0}^{+\infty}V_{1}^{2}\,dV_{1}\cdot\cdot\cdot\int\limits_{0}^{+\infty}V_{N}^{2}\,dV_{N}\,\exp\left\{\frac{8\sin^{2}\frac{\alpha}{2}}{\sigma^{2}}
V_{N} \right\}\,\mathcal {E}_{\sigma \sqrt{\Delta t_{1}}}\left(\frac{\Delta t_{1}}{\Delta x_{1}}V_{N},\,\frac{\Delta t_{1}}{\Delta x_{1}}V_{1}\right)
$$
\begin{equation}
   \label{GN1}
\times
\mathcal {E}_{\sigma \sqrt{\Delta t_{2}}}\left(\frac{\Delta t_{2}}{\Delta x_{2}}V_{1},\,\frac{\Delta t_{2}}{\Delta x_{2}}V_{2}\right)\cdot\cdot\cdot
\mathcal {E}_{\sigma \sqrt{\Delta t_{N}}}\left(\frac{\Delta t_{N}}{\Delta x_{N}}V_{N-1},\,\frac{\Delta t_{N}}{\Delta x_{N}}V_{N}\right)\,.
\end{equation}

Now we can perform functional integration in the functions $\mathcal {E}\,,$ and obtain
 $$
 G^{\alpha}_{N}\left(0,\,t_{1},...t_{N-1} \right)=   \left(\frac{2}{\pi} \right)^{\frac{3}{2}N} \frac{ \sigma^{N}}{ \sqrt{ \Delta t_{1}\cdot\cdot\cdot\Delta t_{N} } }
\ \exp\left\{ \frac{2\pi^{2}}{\sigma^{2}}\left[\frac{1}{ \Delta t_{1}}+ \cdot\cdot\cdot +\frac{1}{\Delta t_{N}} \right]\right\}
 $$
 $$
\times\int\limits_{0}^{1}\chi_{\alpha}(x_{N-1})\,dx_{N-1} \int\limits_{0}^{x_{N-1}}\chi_{\alpha}(x_{N-2})\,dx_{N-2}\cdot\cdot\cdot\int\limits_{0}^{x_{2}}\chi_{\alpha}(x_{1})\,dx_{1}\,\frac{1}{\left(\Delta x_{1}\cdot\cdot\cdot\Delta x_{N} \right)^{2}}
$$
$$
\times\int\limits_{0}^{+\infty}dU_{1}\cdot\cdot\cdot\int\limits_{0}^{+\infty}dU_{N}\ \int\limits_{0}^{+\infty} d\tau_{1}\cdot\cdot\cdot\int\limits_{0}^{+\infty} d\tau_{N}    \ U_{1}\cdot\cdot\cdot U_{N}  \
\exp\left\{8\sin^{2}\frac{\alpha}{2}
U_{N} \right\}
$$
$$
\times\,\exp\left\{-\frac{2}{\Delta x_{1}}\left(U_{N}+U_{1} \right) -  \frac{2}{\Delta x_{2}}\left(U_{1}+U_{2} \right)-...-\frac{2}{\Delta x_{N}}\left(U_{N-1}+U_{N} \right) \right\}
$$
$$
\times\,\exp\left\{-\frac{4}{\Delta x_{1}}\sqrt{U_{N}U_{1} }\cosh \tau_{1} -  \frac{4}{\Delta x_{2}}\sqrt{U_{1}U_{2}}\cosh\tau_{2}-...-\frac{4}{\Delta x_{N}}\sqrt{U_{N-1}U_{N} }\cosh\tau_{N} \right\}
$$
\begin{equation}
   \label{GN2}
\times\,\exp\left\{-\frac{2\tau_{1}^{2}}{\sigma^{2}\Delta t_{1}} \right\}\,\sin\left(\frac{4\pi\tau_{1}}{\sigma^{2}\Delta t_{1}} \right)\,\sinh \tau_{1}
\cdot\cdot\cdot
\exp\left\{-\frac{2\tau_{N}^{2}}{\sigma^{2}\Delta t_{N}} \right\}\,\sin\left(\frac{4\pi\tau_{N}}{\sigma^{2}\Delta t_{N}} \right)\,\sinh \tau_{N}\,.
\end{equation}

\vspace{0.1cm}

The analysis of the dependence of the function $G_{N}^{\alpha} $ on the regularization parameter $\alpha $ as well as the study of possible relations between
different correlation functions will be given in another paper. Here, we only note that
$G_{N}\left(0,\,t_{1},...t_{N-1} \right)$ is singular if there is a pair of coinciding arguments, that is,
 when some $\Delta t_{n}=0\,,  $ similarly to the equation (\ref{GFfinal}).

\section{Concluding remarks}

In this paper, we propose a new approach to study theories invariant under the infinite-dimensional groups of diffeomorphisms and evaluate functional integrals assigning correlation functions in the Schwarzian theory.

Since the Schwarzian theory appears as a limiting theory of various theoretical models, it is of interest to study how the results obtained above could be used in these models.

Having in mind the higher-dimensional as well as supersymmetric versions of the models leading to the Schwarzian theory, it would be desirable  to generalize  the proposed approach and to make it applicable in the corresponding problems.

\vspace{1cm}

\subsection{ A quasi-invariant measure on the group of diffeomorphisms}

In this appendix, we define the measure on the group of diffeomorphisms and give a schematic proof of its quasi-invariance.

Let
$ Diff^{1}_{+} ([0,1])$ be the group of all continuously differentiable transformations of the interval
 $[0,1]$ preserving the ends,
and
  $ Diff^{3}_{+} ([0,1])$  be the subgroup of the group $ Diff^{1}_{+} ([0,1])$ consisting of all diffeomorphisms of the smoothness $C^{3}\,.$

Denote the space of all continuous functions on the interval  $[0,1]$ with zero value at the left end of the interval  by $\ C_{0} ([0,1])\,.$

Consider the map
$$
A : Diff^{1}_{+} ([0,1]) \to C_0 ([0,1])\,,
$$
where
\begin{equation}
\label{Math1}
\xi(t)=\left(A(\varphi)\right) (t) = \log (\varphi'(t)) - \log (\varphi'(0))\,, \ \ \  \forall t \in [0,1]\,.
\end{equation}

The map $A$ identifies the spaces
$\ Diff^1_+ ([0,1])$ and $\ C_0 ([0,1])\,.$
In this case,
\begin{equation}
\label{Math2}
A^{-1} (\xi) (t)=\frac{\int\limits_{0}^{t}\,e^{\xi(\tau)}d\tau}{\int \limits _{0}^{1}e^{\xi(\tau)}d\tau }\,.
\end{equation}

Let  $w_{\sigma}$ be the Wiener measure with the dispersion $\sigma $ on $C_{0}([0,1])\,.$

Now define the measure $\mu_{\sigma}$ on $Diff^{1}_{+}([0,1])$ by the equation
$\mu_{\sigma} (X)=w_{\sigma} (A(X))$ for any measurable subset $X$  of the space $Diff^{1}_{+}([0,1])\,.$

For every $f \in \ Diff^{3}_{+}([0,1])$  and an arbitrary $\varphi \in \ Diff^{1}_{+} ([0,1])$ define
$$
L_{f} (\varphi) =
f \circ \varphi\,.
$$

A detailed proof of the quasi-invariance of the measure $\mu_{\sigma}$ is given in   \cite{(Shavgulidze2000)}).
Here, we present a scheme of the proof.

Consider
$$
\eta=AL_{f}A^{-1}(\xi)\,.
$$
It is written in the form
\begin{equation}
\label{Math3}
\eta(t)=\xi(t)+h \left( \frac{\int\limits_{0}^{t} e^{\xi(\tau)}d\tau}{\int \limits _{0}^{1}e^{\xi(\tau)}d\tau }\right)\,,
\end{equation}
where $h=A(f)$, that is,
$$
h(t)=\log\left(f'(t)\right)-  \log\left(f'(0)\right)\,.
$$

Note that, if $\varphi\in Diff^{3}_{+} ([0,1])\,,$ then $ h\in C_{0}^{2}([0,1])\,.$

The Jacobian of the map (\ref{Math3}) at the point $\xi$ found in the \cite{(Shavgulidze2000)})
does not depend on
 $\xi$ and is equal to
$\frac{1}{f'(1)}\,.$

For continuously differentiable function $\eta$, we obtain
$$
\int \limits _{0}^{1}\left(\eta'(t)\right)^{2}\,dt =\int \limits _{0}^{1}\left(\xi'(t)\right)^{2}\,dt+W+V\,,
$$
where
$$
W=2\int \limits _{0}^{1}\xi'(t)\,h'\left( \frac{\int\limits_{0}^{t} e^{\xi(\tau)}d\tau}{\int \limits _{0}^{1}e^{\xi(\tau)}d\tau }\right)\, \frac { e^{\xi(t)}} {\int\limits_{0}^{1}
e^{\xi(\tau)}d\tau}\,dt\,,
$$
$$
^{}V=\int \limits _{0}^{1}\left(h'\left( \frac{\int\limits_{0}^{t} e^{\xi(\tau)}d\tau }{\int \limits _{0}^{1}e^{\xi(\tau)}d\tau }\right)\right)^{2}\, \frac { e^{2\xi(t)}} {\left(\int\limits_{0}^{1}
e^{\xi(\tau)}d\tau\right)^{2}}\,dt\,.
$$

Let $P$ be a continuous bounded functional on $C_{0}([0,\,1])\,.$ Then
$$
\int \limits _{C_{0}([0,\,1])}P(\eta)\,w_{\sigma}(d\eta)=
\frac{1}{f'(1)}     \int \limits _{C_{0}([0,\,1])}P\left(AL_{f}A^{-1}(\xi)\right)\exp\left\{-\frac{1}{2\sigma^{2}} \left(W+V \right)\right\}\,w_{\sigma}(d\xi)\,.
$$

Note that if $\xi$  were continuously differentiable, then by the integration by parts we would get
$$
\frac{1}{2}W=h'(1)\,\frac { e^{\xi(1)}} {\int\limits_{0}^{1}
e^{\xi(\tau)}d\tau}-h'(0)\,\frac { 1} {\int\limits_{0}^{1}
e^{\xi(\tau)}d\tau}    - \int \limits _{0}^{1}h''\left( \frac{\int\limits_{0}^{t} e^{\xi(\tau)}d\tau}{\int \limits _{0}^{1}e^{\xi(\tau)}d\tau }\right)\, \frac { e^{2\xi(t)}} {\left(\int\limits_{0}^{1}
e^{\xi(\tau)}d\tau\right)^{2}}\,dt\,.
$$
However, the Wiener process $\xi(t)$ is nonsmooth, and there appear the additional terms (\cite{(McKean)}) that can be evaluated in the discrete version of the theory by the correct passage to the continuous limit (see  (\cite{(Shavgulidze2000)})).

As the result, we have
$$
\int\limits_{C_{0}([0,\,1])} P(\eta)\,w_{\sigma}(d\eta)
$$
$$
=\frac{1}{\sqrt{f'(1)f'(1)}}     \int\limits_{C_{0}([0,\,1])} P\left(AL_{f}A^{-1}\xi\right)
\exp\left\{-\frac{1}{2\sigma^{2}} \left( h'(1)\,\frac { e^{\xi(1)}} {\int\limits_{0}^{1}
e^{\xi(\tau)}d\tau}-h'(0)\,\frac {1} {\int\limits_{0}^{1}
e^{\xi(\tau)}d\tau}\right)\right\}
$$
\begin{equation}
\label{Math4}
\times \exp\left\{\,\frac {1}{\sigma
^{2}}\int\limits_{0}^{1}\mathcal{S}_{f} \left( \frac{\int\limits_{0}^{t} e^{\xi(\tau)}d\tau}{\int \limits _{0}^{1}e^{\xi(\tau)}d\tau }\right)\,  \frac { e^{2\xi(t)}} {\left(\int\limits_{0}^{1}
e^{\xi(\tau)}d\tau\right)^{2}}\,dt                \right\}\,w_{\sigma}(d\xi)\,.
\end{equation}

Note that
$$
\mathcal{S}_{f}(t)=
\left(\frac{f''(t)}{f'(t)}\right)'
-\frac{1}{2}\left(\frac{f''(t)}{f'(t)}\right)^2=h''(t)-\frac{1}{2}(h(t))^{2}\,.
$$

The map $A^{-1}$ gives the equation (\ref{FI1})
$$
  \int \limits _{Diff^{1}([0, 1])}\,F(\varphi)\mu_{\sigma}(d\varphi)=\frac{1}{\sqrt{f'(0)f'(1)}}\int \limits _{Diff^{1}([0, 1])}\,F(f\circ\varphi)
  $$
 $$
  \times \exp\left\{ \frac{1}{\sigma^{2}}\left[   \frac{f''(0)}{f'(0)}\varphi'(0)-  \frac{f'(1)}{f'(1)}\varphi'(1)\right]   +        \frac{1}{\sigma^{2}}\int \limits _{0}^{1}\, \mathcal{S}_{f}\left(\varphi(t)\right)\,\left(\varphi' (t)\right)^{2}\,dt  \right\} \,\mu_{\sigma}(d\varphi)\,.
$$

Thus, the measure $\mu_{\sigma}$ on $Diff^{1}_{+}([0,1])$ is quasi-invariant with respect to the subgroup $Diff^{3}_{+}([0,1])\,:$
\begin{equation}
   \label{Math5}
\mu_\sigma (L_{f}  (X))\,=\,\int\limits_{X}\mathcal{R}_{f}^{\mu}(\varphi)
\,\mu_\sigma (d\varphi)\,,
\end{equation}
where the Radon - Nikodim derivative  has the form  (\ref{RadNik})
$$
\mathcal{R}^{\mu}_{f}(\varphi )\equiv\frac{d\mu_{\sigma}^{f}}{d\mu_{\sigma}}(\varphi)=\frac{1}{\sqrt{f'(0)f'(1)}}
$$
$$
  \times \exp\left\{ \frac{1}{\sigma^{2}}\left[   \frac{f''(0)}{f'(0)}\varphi'(0)-  \frac{f''(1)}{f'(1)}\varphi'(1)\right]   +        \frac{1}{\sigma^{2}}\int \limits _{0}^{1}\, \mathcal{S}_{f}\left(\varphi(t)\right)\,\left(\varphi' (t)\right)^{2}\,dt  \right\} \,.
$$

\subsection{Proof of the basic formula}

To evaluate the basic Wiener integral (\ref{FormulaX}) we use the quasi-invariance of the Wiener measure
under the action of the operator $K_{f}\equiv A\,L_{f}\,A^{-1}\,:$
$$
\left(K_{f}\,\xi\right)(t)=\xi(t)+\log\left\{f'\left(\frac{\int\limits_{0}^{t} e^{\xi(\tau)}d\tau}{\int \limits _{0}^{1}e^{\xi(\tau)}d\tau } \right) \right\}-\log\left\{f'(0)\right\}\,.
$$
The Radon-Nikodim derivative is
$$
\frac{dw_{\sigma}^{f}}{dw_{\sigma}}(\xi)
=\frac{1}{\sqrt{f'(0)f'(1)}}
\exp\left\{ \frac{1}{\sigma^{2}}\left[   \frac{f''(0)}{f'(0)}-  \frac{f''(1)}{f'(1)}e^{\xi(1)}\right]  \frac{1}{\int \limits _{0}^{1}e^{\xi(\tau)}d\tau }\right\}
$$
$$
\times \exp \left\{\frac{1}{\sigma^{2}}\int \limits _{0}^{1}\, \mathcal{S}_{f}\left(\frac{ \int\limits_{0}^{t} e^{\xi(\tau)}d\tau}{\int \limits _{0}^{1}e^{\xi(\tau)}d\tau } \right)\,\frac{ e^{2\xi(t)}}{\left(\int \limits _{0}^{1}e^{\xi(\tau)}d\tau \right)^{2}}\,dt  \right\} \,,
$$
where $w_{\sigma}^{f}(X)=w_{\sigma}(K_{f}X )\,.$

Now consider the special transformation
$$
f=g_{\beta}(t)=\frac{(\beta+1)t}{\beta t+1}
$$
with
$$
g'_{\beta}(t)=\frac{\beta+1}{(\beta t+1)^{2}}\,,\ \ \ \ \ \ \ g''_{\beta}(t)=-\frac{2(\beta+1)\beta}{(\beta t+1)^{3}}\,,\ \ \ \ \  \ \ \mathcal{S}_{g_{\beta}}(t)=0\,.
$$

In this case, the Radon-Nikodim derivative has the form
\begin{equation}
   \label{RNbeta}
\frac{dw_{\sigma}^{g_{\beta}}}{dw_{\sigma}}(\xi)
=\exp\left\{ \frac{-2\beta}{\sigma^{2}}\left[  1-  \frac{e^{\xi(1)}}{\beta+1 }\right]  \frac{1}{\int \limits _{0}^{1}e^{\xi(\tau)}d\tau }\right\}\,,
\end{equation}
and
$$
\int\limits_{C_{0}([0,\,1])} \delta\left(\xi(1)-x\right)\,w^{g_{\beta}}_{\sigma}(d\xi)
$$
$$
=\int\limits_{C_{0}([0,\,1])} \delta\left(\xi(1)-x\right)\exp\left\{ \frac{-2\beta}{\sigma^{2}}\left[  1-  \frac{e^{\xi(1)}}{\beta+1 }\right]  \frac{1}{\int \limits _{0}^{1}e^{\xi(\tau)}d\tau }\right\}w_{\sigma}(d\xi)\,.
$$

At the same time,
$$
\int\limits_{C_{0}([0,\,1])} \delta\left(\xi(1)-x\right)\,w^{g_{\beta}}_{\sigma}(d\xi)=\int\limits_{C_{0}([0,\,1])} \delta\left(\eta(1)+2\log(\beta+1)-x\right)w_{\sigma}(d\eta)
$$
$$
=\frac{1}{\sqrt{2\pi}\sigma}\exp \left\{-\frac{1}{2\sigma ^{2}}\left( x-2\log (\beta +1)\,\right)^{2} \right\}\,.
$$
Here, we have used the equation
$$
\int\limits_{C_{0}([0,\,1])} \delta\left(\xi(1)-x\right)\,w_{\sigma}(d\xi)=\frac{1}{\sqrt{2\pi}\sigma}\,\exp \left\{-\frac{1}{2\sigma ^{2}} x^{2} \right\}\,,
$$
and the following relation:
$$
\eta(1)=\left(K_{g_{\beta}}\xi\right)(1)=\xi(1)+\log g'_{\beta}(1)-\log g'_{\beta}(0)=\xi(1)-2\log (\beta +1)\,,
$$
or
$$
\xi(1)=\eta(1)+2\log (\beta +1)\,.
$$

Therefore, the basic formula
$$
\int\limits_{C_{0}([0,\,1])} \delta\left(\xi(1)-x\right)\,\exp\left\{ \frac{-2\beta}{\sigma^{2}}\left[  1-  \frac{e^{\xi(1)}}{\beta+1 }\right]  \frac{1}{\int \limits _{0}^{1}e^{\xi(\tau)}d\tau }\right\}\,w_{\sigma}(d\xi)
$$
\begin{equation}
   \label{XFormula}
  =\frac{1}{\sqrt{2\pi}\sigma}\,\exp \left\{-\frac{1}{2\sigma ^{2}}\left( x-2\log (\beta +1)\,\right)^{2} \right\}
\end{equation}
is proven.

\subsection{Properties of the function $\mathcal {E}_{\sigma}(u,\,v) $ }

To study the properties of the function $\mathcal {E}_{\sigma}(u,\,v) $ given by the equation (\ref{E})
$$
\mathcal {E}_{\sigma}(u,\,v)=\int\limits_{Diff^{1} ([0,1]) }\,\delta\left(\,\varphi'(0)-u \,\right)\,\delta\left(\,\varphi'(1)-v \, \right)\,\mu_{\sigma}(d\varphi)
$$
consider the functions $\varphi,\, \psi\in Diff^{1} ([0,1])$ connected by the diffeomorphism $\,g_{\lambda}\,:$
\begin{equation}
   \label{glambda}
\varphi(t)=g_{\lambda}(\psi(t))\,, \ \ \ \ \ g_{\lambda}(\tau)=\frac{(\lambda-1)\tau}{\lambda -\tau}\,, \ \ \lambda>1\,,\ \ \ g_{\lambda}\in Diff^{3}([0,\,1])\,.
\end{equation}

In this case, the equation (\ref{FI1}) gives
$$
  \mathcal {E}_{\sigma}(u,\,v)=\int\limits_{Diff^{1} ([0,1]) }\,\delta\left(\,g'_{\lambda}(0)\psi'(0)-u \,\right)\,\delta\left(\,g'_{\lambda}(1)\psi'(1)-v\, \right)
$$
$$
  \times \exp\left\{ \frac{1}{\sigma^{2}}\left[   \frac{g''(0)}{g'(0)}\psi'(0)-  \frac{g''(1)}{g'(1)}\psi'(1)\right]  \right\} \,\mu_{\sigma}(d\psi)
$$
\begin{equation}
   \label{Epsi}
 =\int\limits_{Diff^{1} ([0,1]) }\delta\left(\psi'(0)-\frac{\lambda}{\lambda -1}u \right)\delta\left(\psi'(1)-\frac{\lambda -1}{\lambda}v \right)\exp\left\{ \frac{2}{\sigma^{2}}\left(   \frac{u}{\lambda -1}-  \frac{v}{\lambda}\right)  \right\} \mu_{\sigma}(d\psi)\,.
\end{equation}

Therefore, we have
\begin{equation}
   \label{Elambda}
 \mathcal {E}_{\sigma}(u,\,v)=\mathcal {E}_{\sigma}\left(\frac{\lambda}{\lambda -1}u,\,\frac{\lambda -1}{\lambda}v\right)\,\exp\left\{ \frac{2}{\sigma^{2}}\left(   \frac{u}{\lambda -1}-  \frac{v}{\lambda}\right)  \right\}\,.
\end{equation}
In particular, for
$$
\frac{1}{\lambda}=1-\sqrt{\frac{u}{v}}\,,
$$
the above equation has the symmetric form (\ref{Esymm}):
$$
\mathcal {E}_{\sigma }\left(u,\,v\right)=\mathcal {E}_{\sigma }\left(\sqrt{uv},\,\sqrt{uv}\right)\,\exp \left\{ -\frac{2}{\sigma^{2}}\left(\sqrt{u}-\sqrt{v}\right)^{2} \right\}\,.
$$

Now we perform the functional integration in  $\mathcal {E}_{\sigma}(u,\,u)\,. $ First we write it in the form of the integrals over the Wiener measure
$$
\mathcal {E}_{\sigma}(u,\,u)=\int\limits_{Diff^{1} ([0,1]) }\delta\left(\varphi '(0)-u \right)\,\delta\left(\varphi '(1)-u \right)\,\mu_{\sigma}(d\varphi)
$$
$$
=\int\limits_{Diff^{1} ([0,1]) }\delta\left(\,\varphi '(0)-u\, \right)\,\delta\left(\,\varphi '(1)- \varphi '(0)\, \right)\,\mu_{\sigma}(d\varphi)
$$
$$
=\int\limits_{C_{0} ([0,\,1]) }\delta \left(\frac{1}{\int \limits _{0}^{1}e^{\xi(\tau)}d\tau }-u \right)
 \delta \left(u\,\left[\,e^{\xi(1)} - 1\, \right ]\right)\,w_{\sigma}(d\xi)
$$
\begin{equation}
\label{EWiener}
=\frac{1}{u}\int\limits_{C_{0} ([0,\,1]) }\delta \left(\frac{1}{\int \limits _{0}^{1}e^{\xi(\tau)}d\tau }-u \right)
 \delta\left(\xi(1)\right)\,w_{\sigma}(d\xi)\,.
\end{equation}
Taking the Fourier transform of the first $\delta-$function in (\ref{EWiener}), we get
\begin{equation}
\label{EWiener2}
\mathcal {E}_{\sigma}(u,\,u)=\frac{1}{2\pi\,u}\,\int\limits_{-\infty}^{+\infty}\,e^{i\rho u}d\rho \,\int\limits_{C_{0} ([0,\,1]) }\exp \left\{-i\rho\frac{1}{\int \limits _{0}^{1}e^{\xi(\tau)}d\tau } \right\}
 \delta\left(\xi(1)\right)\,w_{\sigma}(d\xi)\,.
\end{equation}
It is convenient to rewrite the above equation as
$$
\mathcal {E}_{\sigma}(u,\,u)=\frac{1}{u\,\pi\,\sigma^{2}}\,\int\limits_{-\infty}^{+\infty}\,\exp \left\{i u\frac{2r}{\sigma^{2}}\right\}dr
$$
\begin{equation}
\label{EWiener3}
\times\int\limits_{C_{0} ([0,\,1]) }\exp \left\{-i\frac{2r}{\sigma^{2}}\frac{1}{\int \limits _{0}^{1}e^{\xi(\tau)}d\tau } \right\}
 \delta\left(\xi(1)\right)\,w_{\sigma}(d\xi)\,.
\end{equation}

To use the equation (\ref{Formula}), note that a solution of the equation
$$
\frac{\beta^{2}}{\beta +1}=ir
$$
has the form
$$
\beta_{\ast}(r)+1= \left(i\frac{r}{2}+1\right)+\sqrt{\left(i\frac{r}{2}+1\right)^{2}-1}\,.
$$
Due to the identity
$$
\log\left(y+\sqrt{y^{2}-1} \right)=arc\cosh y\,,
$$
the equation (\ref{Formula}) gives
$$
\int\limits_{C_{0} ([0,\,1]) }\exp \left\{-i\frac{2r}{\sigma^{2}}\frac{1}{\int \limits _{0}^{1}e^{\xi(\tau)}d\tau } \right\}
 \delta\left(\xi(1)\right)\,w_{\sigma}(d\xi)
$$
$$
= \frac{1}{\sqrt{2\pi}\sigma}\exp\left\{-\frac{2}{\sigma^{2}} \left(arc\cosh \left( \frac{ir}{2}+1\right) \right)^{2}\right\}\,.
$$

Therefore, the function $\mathcal {E}_{\sigma}(u,\,u) $ is written as
\begin{equation}
\label{Earccosh}
\mathcal {E}_{\sigma}(u,\,u)=\frac{2}{u\,(2\pi)^{\frac{3}{2}}\,\sigma^{3}}\,\int\limits_{-\infty}^{+\infty}\,\exp \left\{i u\frac{2r}{\sigma^{2}}\right\}
\exp\left\{-\frac{2}{\sigma^{2}} \left(arc\cosh \left( \frac{ir}{2}+1\right) \right)^{2}\right\}\,dr \,.
\end{equation}
Having in mind the properties of the function $arc\cosh $, we turn the integration contour with the result
$$
\mathcal {E}_{\sigma}(u,\,u)=\frac{\sqrt{2}}{u\,(2\pi)^{\frac{3}{2}}\,\sigma^{3}}\exp\left\{\frac{2\pi^{2}}{\sigma^{2}} \right\}\int\limits_{4}^{+\infty}\,\exp \left\{- \frac{2ux}{\sigma^{2}}\right\}
$$
\begin{equation}
\label{Earccosh2}
\times
\exp\left\{-\frac{2}{\sigma^{2}} \left(arc\cosh \left( \frac{x-2}{2}\right) \right)^{2}\right\}\sin\left( \frac{4\pi}{\sigma^{2}}arc\cosh \left( \frac{x-2}{2}\right)\right)dx \,.
\end{equation}
After the substitution
$$
\tau=arc\cosh \left( \frac{x-2}{2}\right)\,,\ \ \ \ \ x=2+2\cosh \tau\,,
$$
the function $\mathcal {E}_{\sigma}(u,\,u) $ takes the form (\ref{Eint})
$$
\mathcal {E}_{\sigma }\left(u,\,u\right)=\frac{2\sqrt{2}}{u\,\pi^{\frac{3}{2}}\sigma^{3}}\,\exp\left\{\frac{2\pi^{2}-4\,u}{\sigma^{2}} \right\}
$$
$$
\times \int\limits_{0}^{+\infty}\,\exp\left\{-\frac{2}{\sigma^{2}}\left(2\,u\,\cosh\tau+\tau^{2}\right) \right\}\,\sin\left(\frac{4\pi\tau}{\sigma^{2}} \right)\,\sinh(\tau)\,d\tau \,.
$$

The forms of the functions $\mathcal {E}_{\sigma}(x,\,x) $ and $\mathcal {E}_{\sigma}(x,\,y)$  at $\sigma=1 $ are presented at Fig.2 and at Fig.3 respectively.

\begin{figure}
\includegraphics [height=0.3\textwidth]{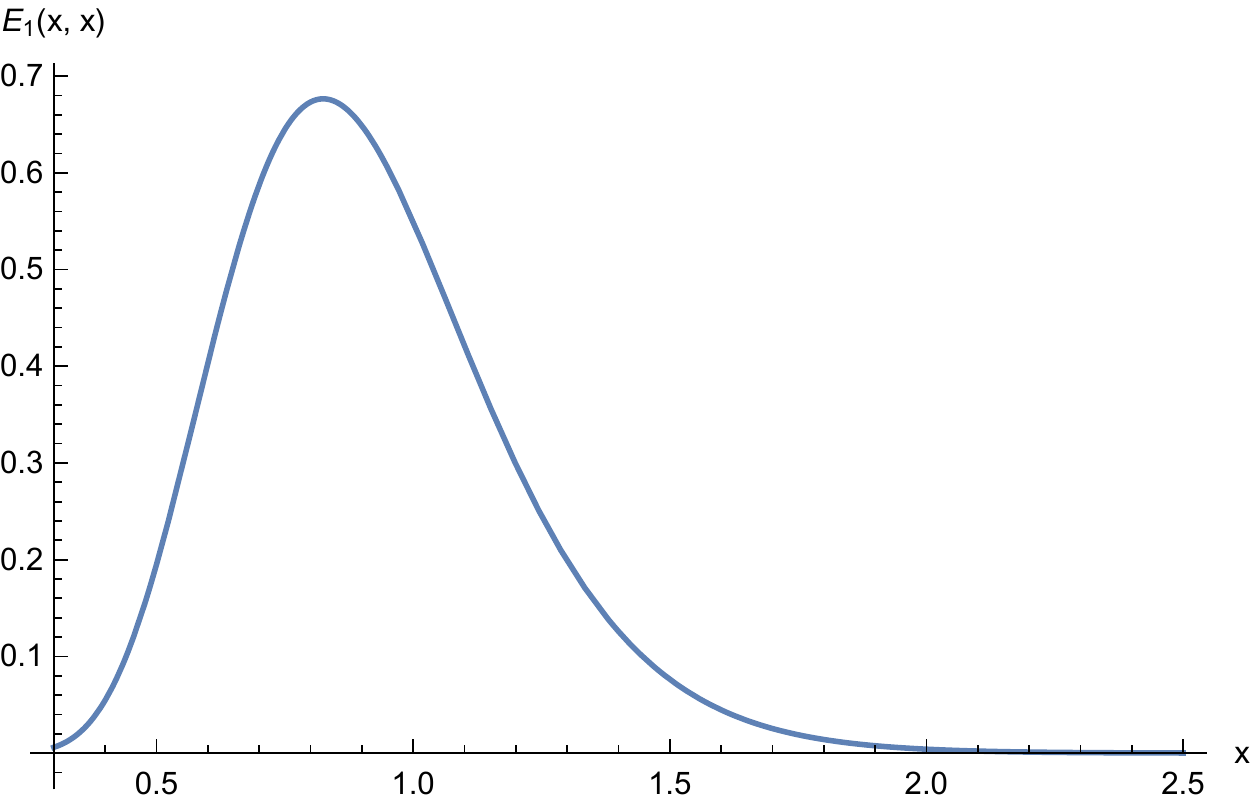}
\caption{The form of the symmetric function  $\mathcal {E}_{1}(x,\,x)\,. $}
\end{figure}

\begin{figure}
\includegraphics [height=0.3\textwidth]{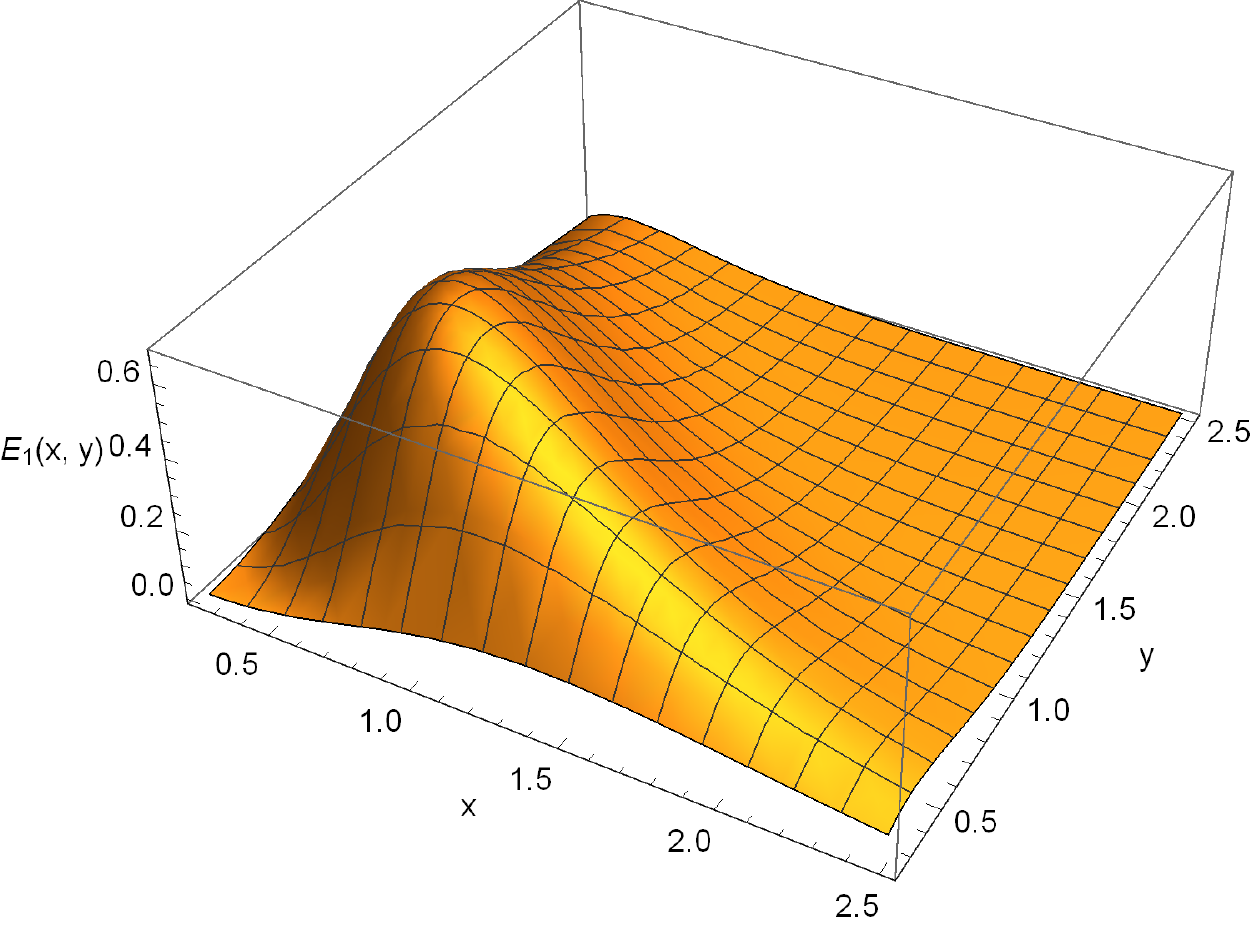}
\caption{The form of the function  $\mathcal {E}_{\sigma}(x,\,y)\,.$ }
\end{figure}

\subsection{ Integration in $SL(2, \textbf{R})$}

In this appendix, we evaluate the asymptotic form (at $\pi-\alpha\rightarrow 0$) of the regularized integrals over the Haar measure $d\nu $ on the group $SL(2, \textbf{R})\,.$
These integrals are used to normalize the corresponding functional integrals over the group $Diff^{1}$ considered in this paper. We are interesting in the asymptotic form of the following integrals:
\begin{equation}
   \label{V2}
V^{\alpha}_{SL(2, \textbf{R})}=\int \limits _{SL(2,\textbf{R})}\exp \left\{-\lambda\,\int \limits _{0}^{1}\dot{\varphi}^{2}(t)dt  \right\}  d\nu\,,
\end{equation}
\begin{equation}
   \label{Fi2}
\Phi^{\alpha}_{SL(2, \textbf{R})}=\int \limits _{SL(2,\textbf{R})}\dot{\varphi}(0)\,\exp \left\{-\lambda\,\int \limits _{0}^{1}\dot{\varphi}^{2}(t)dt  \right\}  d\nu\,,
\end{equation}
\begin{equation}
   \label{G2SL2R}
G^{\alpha}_{2;SL(2, \textbf{R})}=\int \limits _{SL(2,\textbf{R})}\dot{\varphi}(t)\dot{\varphi}(0)\,\exp \left\{-\lambda\,\int \limits _{0}^{1}\dot{\varphi}^{2}(t)dt  \right\}  d\nu\,.
\end{equation}
Here,
$$
\lambda=  \frac{2\left[ \pi^{2}-\alpha^{2}\right]}{\sigma^{2}}\,.
$$

To perform the integration over the group $SL(2,\textbf{R})$  we choose the representation \cite{(Lang)}
\begin{equation}
   \label{Repres}
  \varphi_{z}(t)=-\frac{i}{2\pi}\log\frac{e^{i2\pi t}+z}{\bar{z}\,e^{i2\pi t}+1}\,, \ \ \ z=\rho e^{i\theta}\,, \ \ \ \rho < 1  \,.
\end{equation}
In this case, the Haar measure is \cite{(Lang)}
\begin{equation}
   \label{Haar}
  \nu(dz)=\frac{2dz \,d\bar{z}}{\left(1-z\bar{z} \right)^{2}}=\frac{4\rho d\rho \,d\theta}{\left(1-\rho^{2} \right)^{2}} \,.
\end{equation}

To evaluate the integral
\begin{equation}
   \label{dt}
 I=\int \limits _{0}^{1}\dot{\varphi}_{z}^{2}(t)dt =\left(1-|z|^{2} \right)^{2} \,\int \limits _{0}^{1}\frac{dt}{\left( e^{i2\pi t}+z \right)^{2}\left(e^{-i2\pi t}+\bar{z} \right)^{2}}
\,,
\end{equation}
note that, due to the periodicity, it does not depend on $\theta \,.$ Therefore, we can assume $z$ to be real $z=\rho>0\, ,\ \ \rho<1\,.$
After the substitution $w=\rho\,\exp\{i2\pi t\}\,,$ the integral $I$ transforms into the contour integral
$$
 I=\frac{1}{2\pi i}\left(1-\rho^{2} \right)^{2} \,\oint \limits _{|w|=\rho<1}\frac{w\,dw}{\left( w+\rho^{2} \right)^{2}\left(w+1 \right)^{2}}
$$
 \begin{equation}
   \label{contour}
 =\left(1-\rho^{2} \right)^{2} \,Res  _{\{w=-\rho^{2}\}}\,\frac{w}{\left( w+\rho^{2} \right)^{2}\left(w+1 \right)^{2}}
 =-1+\frac{2}{1-\rho^{2}}=-1+\frac{2}{1-z\bar{z}}\,.
\end{equation}

And the regularized volume of the group (\ref{V2}) has the form
$$
V^{\alpha} _{SL(2, \textbf{R})}=e^{\lambda}\int \limits_{|z|<1}\exp \left\{-\lambda \frac{2}{(1-z\bar{z})} \right\}\frac{2dz d\bar{z}}{\left(1-z\bar{z} \right)^{2}}
=e^{\lambda}\int \limits _{0}^{1}\exp\left\{-\lambda\frac{2}{(1-\rho^{2})} \right\}
\frac{8\pi \rho d\rho }{\left(1-\rho^{2} \right)^{2}}=\frac{2\pi}{\lambda}\,.
$$
Thus, at $\alpha\rightarrow\pi-0\,, $
\begin{equation}
   \label{Vas}
V^{As} _{SL(2, \textbf{R})}= \frac{\sigma^{2}}{2(\pi-\alpha)}\,.
\end{equation}

To find the second integral
$$
\Phi^{\alpha} _{SL(2, \textbf{R})}=e^{\lambda}\,\int \limits_{|z|<1}\exp \left\{-\lambda \frac{2}{(1-z\bar{z})} \right\}\frac{1}{(1+z)(1+\bar{z})}\frac{2dz \,d\bar{z}}{\left(1-z\bar{z} \right)}\,,
$$
it is convenient to use the following representation of the complex variable $z$:
\begin{equation}
   \label{z2}
z=-1+\varrho e^{i\vartheta}\,,
\end{equation}
and write (\ref{Fi2}) in the form
\begin{equation}
   \label{Fi3}
\Phi^{\alpha}_{SL(2, \textbf{R})}=2\,e^{\lambda}\,\int \limits _{-\frac{\pi}{2}}^{+\frac{\pi}{2}}d\vartheta\,\int \limits _{0}^{2\cos \vartheta}\,\exp \left\{-\frac{2\lambda}{\varrho\,(2\cos \vartheta-\varrho)} \right\} \frac{d\varrho}{\varrho ^{2}\,(2\cos \vartheta-\varrho)}\,.
\end{equation}
Note that in terms of $\varrho$ and $\vartheta$, the integral in (\ref{V2}) looks like
\begin{equation}
   \label{V3}
V^{\alpha}_{SL(2, \textbf{R})}=2\,e^{\lambda}\,\int \limits _{-\frac{\pi}{2}}^{+\frac{\pi}{2}}d\vartheta\,\int \limits _{0}^{2\cos \vartheta}\,\exp \left\{-\frac{2\lambda}{\varrho\,(2\cos \vartheta-\varrho)} \right\} \frac{d\varrho}{\varrho\,(2\cos \vartheta-\varrho)^{2}}\,.
\end{equation}
The substitution
$$\tilde{\varrho}=2\cos \vartheta-\varrho
$$
transforms the integral (\ref{Fi3}) into the integral (\ref{V3}).

Thus we have
\begin{equation}
   \label{Fias}
\Phi^{As} _{SL(2, \textbf{R})}= \frac{\sigma^{2}}{2(\pi-\alpha)}\,.
\end{equation}

Now we find the asymptotic form of the integral
$$
G^{\alpha} _{2;\,SL(2, \textbf{R})}\left(0,\,\frac{1}{2}\right)=e^{\lambda}\int \limits_{|z|<1}\exp \left\{-\lambda \frac{2}{(1-z\bar{z})} \right\}\frac{2dz \,d\bar{z}}{(1+z)(1+\bar{z})(1-z)(1-\bar{z})}
$$
\begin{equation}
   \label{G2as1}
=e^{\lambda}\int \limits_{0}^{2\pi}d\theta\,\int \limits_{0}^{1}\exp \left\{-\lambda \frac{2}{(1-\rho^{2})} \right\}\frac{2d\rho^{2} } {(1-\rho^{2})^{2}+4\rho^{2}\sin^{2}\theta}\,.
\end{equation}

The method widely used to study the asymptotic behavior of Feynman diagrams (see, e.g.,   \cite{(Smirn)} and refs. therein) is very helpful here.
Namely, we consider the Mellin transform of the integrand and rewrite (\ref{G2as1}) as follows:
$$
G^{\alpha} _{2;\,SL(2, \textbf{R})}\left(0,\,\frac{1}{2}\right)=e^{\lambda}\int \limits_{a_{0}-i\infty}^{a_{0}+i\infty}da\,\Gamma(a)\Gamma (1-a)\int \limits_{0}^{2\pi}d\theta \left(sin^{2}\theta \right)^{-a}
$$
\begin{equation}
   \label{G2as2}
\times\int \limits_{0}^{1}\exp \left\{-\lambda \frac{2}{(1-\rho^{2})} \right\}\frac{2d\rho^{2} } {4^{a}(1-\rho^{2})^{2(1-a)}\,\rho^{2a}}\,,\ \ \ 0<a_{0}<1\,.
\end{equation}

After the substitution $x=(1-\rho^{2})^{-1}\,,$
the integrals over $\theta$ and over $x$  are the table integrals (see, e.g, \cite{(Prud)} n. 2.5.3.2 and n. 2.3.6.7):
$$
\int \limits_{0}^{\frac{\pi}{2}}\sin^{-2a}\theta\,d\theta =\frac{1}{2}\frac{\Gamma\left(\frac{1}{2}-a \right)\Gamma\left(\frac{1}{2} \right)}{\Gamma\left(1-a \right) }\,,
$$
$$
\int \limits_{1}^{+\infty}x^{-a}(x-1)^{-a}\,e^{-2\lambda x}\,dx
=\frac{e^{-\lambda}\sqrt{\pi}\Gamma (1-a)}{2\sin\left(\left[\frac{1}{2}-a \right] \pi\right)}\,(2\lambda)^{-\frac{1}{2}+a}\left[I_{-\frac{1}{2}+a}(\lambda)- I_{\frac{1}{2}-a}(\lambda)\right]\,.
$$
Here, $I_{\nu}(\lambda)$ is the modified Bessel function
$$
I_{\nu}(\lambda)=\sum \limits_{k=0}^{\infty}\frac{\lambda^{k+\nu}}{k!2^{k+\nu}\Gamma (k+\nu+1)}\,.
$$
Note that the first term in the series $(k=0)$ is the leading one at $\lambda\rightarrow 0\,. $ As the result, the integral is reduced to
\begin{equation}
   \label{G2as3}
G^{\alpha} _{2;\,SL(2, \textbf{R})}\left(0,\,\frac{1}{2}\right)=\pi\,\int \limits_{a_{0}-i\infty}^{a_{0}+i\infty}da\,\frac{\Gamma(a)\Gamma (1-a)}{4^{a}\Gamma \left( \frac{1}{2}+a\right)}\,\frac{\Gamma \left( \frac{1}{2}-a\right)}{\sin\left(\left[\frac{1}{2}-a \right] \pi\right)}\,\lambda^{2a-1}\,.
\end{equation}

Then we close the integration contour in the right-hand half-plane and note that the integral over the infinite half-circle equals to zero.

Now the leading asymptotics of the function $G^{\alpha} _{2;\,SL(2, \textbf{R})}\left(0,\,\frac{1}{2}\right)$ is given by the residue at the pole of the integrand inside the contour with the minimal value of $a\,.$ For (\ref{G2as3}), it is the pole at $a=\frac{1}{2}\,$ and
\begin{equation}
   \label{G2as4}
G^{As} _{2;\,SL(2, \textbf{R})}\left(0,\,\frac{1}{2}\right)=-\pi\,\log(\pi-\alpha)\,.
\end{equation}


\begin{thebibliography}{7}

\bibitem{(Kit1)}
A. Kitaev, \emph{Hidden correlations in the Hawking radiaion and thermal noise, Talk at KITP},
http://online.kitp.ucsb.edu/online/joint98/kitaev/, Febrary, 2015.

\bibitem{(Kit2)}
A. Kitaev, \emph{A simple model of quantum holography, Talks at KITP},
http://online.kitp.ucsb.edu/online/entangled15/kitaev/ and http://online.kitp.ucsb.edu/online/entangled15/kitaev2/, April and May, 2015.

\bibitem{(MS)}
J. Maldacena and D. Stanford, \emph{Remarks on the Sachdev-Ye-Kitaev model, Phys. Rev.} \textbf{D 94}(2016) 106002, arXiv:1604.07818 [hep-th].

\bibitem{(GR)}
D. J. Gross and V. Rosenhaus, \emph{A Generalization of Sachdev-Ye-Kitaev, JHEP } \textbf{ 02} (2017) 093, arXiv:1610.01569 [hep-th].

\bibitem{(MNW)}
G. Mandal, P. Nayak and S. R. Wadia, \emph{Coadjoint orbit action of Virasoro group and two-dimentional quantum gravity dual SYK/tensor models}, arXiv:1702.04266 [hep-th].

\bibitem{(SW)}
D. Stanford and E. Witten, \emph{Fermionic Localization of the Schwazian Theory, JHEP} \textbf{10} (2017) 008, arXiv:1703.04612 [hep-th].

\bibitem{(KitSuh)}
A. Kitaev and S. J. Suh, \emph{The soft mode in the Sachdev-Ye-Kitaev model and its gravity dual}, arXiv:1711.08467v2 [hep-th].

\bibitem{(Mertens)}
T. G. Mertens,  \emph{The Schwarzian Theory - Origins}, arXiv:1801.09605v2 [hep-th].

\bibitem {(Kuo)} Hui-Hsiung Kuo, \emph{Gaussian Measures in Banach Spaces, Springer, Berlin-Heidelberg-NY,} 1975.

\bibitem{(BAK)}
D. Bagrets, A. Altland and A. Kamenev, \emph{Sachdev-Ye-Kitaev Model as Liouville Quantum Mechanics, Nucl. Phys.} \textbf{ B 911} (2016) 191, arXiv:1607.00694 [cond-mat.str-el].

\bibitem{(BAK2)}
D. Bagrets, A. Altland and A. Kamenev, \emph{Power-law out of time order correlation functions in the SYK model, Nucl. Phys.} \textbf{B 921} (2017) 727, arXiv:1702.08902v1 [cond-mat.str-el].

\bibitem{(MTV)}
T. G. Mertens, G. J. Turiaci and H. L. Verlinde, \emph{Solving the Schwarzian via the Conformal Bootstrap, JHEP} \textbf{08} (2017) 136, arXiv:1705.08408v3 [hep-th].

\bibitem{(GR2)}
D. J. Gross and V. Rosenhaus, \emph{All point correlation functions in SYK}, arXiv:1710.08113 [hep-th].

\bibitem{(BShExact)}
V. V. Belokurov and E. T. Shavgulidze,  \emph{Exact solution of the Schwarzian theory, Phys. Rev.} \textbf{D 96}(2017) 101701(R) , arXiv:1705.02405 [hep-th].

\bibitem {(Neretin1996)} Yu. A. Neretin, \emph{Categories of Symmetries and Infinite-Dimensional Groups, Clarendon Press, Oxford}, 1996.

\bibitem {(Shepp)} L. A. Shepp, \emph{ Radon - Nikodym Derivatives of Gaussian measures, Ann. Math. Statistics,} \textbf{37} (1966) 321.

\bibitem{(BSh)}
V. V. Belokurov and E. T. Shavgulidze, \emph{Extraordinary Properties of Functional Integrals and Groups of Diffeomorphisms, Physics of Particles and Nuclei} \textbf{48} (2017) 267 [\emph{Fizika Elementarnych Chastits @ Atomnogo Yadra } \textbf {48} (2017) 194].

\bibitem {(Shavgulidze2018)} E. T. Shavgulidze, \emph{Amenability of Discrete Subgroups of Groups of Diffeomorphisms, to be published.}

\bibitem {(Shavgulidze1978)} E. T. Shavgulidze,  \emph{An example of a measure quasi-invariant with respect to the action of a group of diffeomorphisms of the circle, Functional Analysis and Applications} \textbf{12} (1978) 55.

\bibitem {(Shavgulidze1988)} E. T. Shavgulidze, \emph{A measure quasi-invariant with respect to the action of a group of diffeomorphisms of a finite-dimensional manifold, Soviet Mathematical Doklady} \textbf{38} (1988) 622.

\bibitem {(Shavgulidze2000)} E. T. Shavgulidze, \emph{Some Properties of Quasi-Invariant
Measures on Groups of Diffeomorphisms of the Circle, Russian
Journ. Math. Phys.} \textbf{7} (2000) 464.

\bibitem{(Prud)}
A. P. Prudnikov, Yu. A. Brychkov, O. I. Marichev, \emph{Integrals and Series. V. 1. Elementary Functions, Gordon Breach: New York-London}, 1986.

\bibitem {(McKean)} H. P. McKean, \emph{ Stochastic Integrals, Academic Press, NY-London}, 1969.

\bibitem{(Lang)}
S. Lang, "$SL_{2}(\textbf{R})$". \emph{Addison-Wesley Publishing. }1975.

\bibitem{(Smirn)}
V. A. Smirnov,  \emph{Feynman Integrals Calculus, Springer: Berlin-Heidelberg-New york }, 2006.










\end{thebibliography}
\end{document}